# A Compression-Complexity Measure of Integrated Information

Mohit Virmani[*◑], Nithin Nagaraj[◑]

Consciousness Studies Programme, National Institute of Advanced Studies, IISc Campus, Bengaluru, Karanataka, India

◑These authors contributed equally to this work.
* mohitvirmani11@gmail.com

## Abstract

Quantifying integrated information is a leading approach towards building a fundamental theory of consciousness. Integrated Information Theory (IIT) has gained attention in this regard due to its theoretically strong framework. However, it faces some limitations such as current state dependence, computationally expensive and inability to be applied to real brain data. On the other hand, Perturbational Complexity Index (PCI) is a clinical measure for distinguishing different levels of consciousness. Though PCI claims to capture the functional differentiation and integration in brain networks (similar to IIT), its link to integrated information theories is rather weak. Inspired by these two approaches, we propose a new measure - $\Phi^C$ using a novel compression-complexity perspective that serves as a bridge between the two, for the first time. $\Phi^C$ is founded on the principles of lossless data compression based complexity measures which characterize the dynamical complexity of brain networks. $\Phi^C$ exhibits following salient innovations: (i) mathematically well bounded, (ii) negligible current state dependence unlike $\Phi$, (iii) integrated information measured as compression-complexity rather than as an infotheoretic quantity, and (iv) faster to compute since number of atomic bipartitions scales linearly with the number of nodes of the network, thus avoiding combinatorial explosion. Our computer simulations show that $\Phi^C$ has similar hierarchy to $< \Phi >$ for several multiple-node networks and it demonstrates a rich interplay between differentiation, integration and entropy of the nodes of a network. $\Phi^C$ is a promising heuristic measure to characterize the quantity of integrated information (and hence a measure of quantity of consciousness) in larger networks like human brain and provides an opportunity to test the predictions of brain complexity on real neural data.

## Author Summary

Integrated Information Theory (IIT) has recently gained a lot of attention as a promising candidate for a scientific theory of consciousness. IIT is a theoretical approach that measures the capacity of brain networks to differentiate between a large number of experiences and yet act as an integrated system. However, IIT has several limitations such as sensitivity to current states of the network, computationally very expensive and hence inapplicable as a clinically useful measure. At the other extreme, a clinical measure for distinguishing levels of consciousness, known as Perturbational Complexity Index (PCI) has been proposed recently. However, PCI doesn't have an established theoretical link to information integration theories. Inspired by IIT and PCI, we introduce the idea of compression-complexity and propose a novel measure of integrated information. Current state independence, ease of computation, robustness and applicability to time series data are some of the innovations of our measure which pave the way for applications to neurophysiological measurements and data from complex networks (biological or otherwise).

## Introduction

Consciousness is our "subjective experience", which is unique and peculiar in its own sense such as a feeling of pain, perceived sensation of color or in more general sense the experience felt by any organism i.e. "What's it like to be?" [1]. Consciousness is hard enough to be defined in words but easiest to be accepted, as it is something rather than nothing, which each of us is experiencing right now. Understanding consciousness and its measures are even more important than before, because of the upsurge of smart learning algorithms [2,3], which makes us doubt if machines possess consciousness or not. The problem of measuring consciousness is difficult because of the presence of different levels of conscious experience [4] and first person reports of consciousness might not be accurate. It has also been suggested that we need a mix of theoretical and practical approaches to be able to define and measure the quantity of consciousness [5, 6].

On the basis of various scientific theories, different measures of consciousness are suggested in the literature - both on behavioural and neurophysiological basis [4]. The idea that consciousness is the result of a balance between functional integration and differentiation in thalamocortical networks, or brain complexity, has gained recent popularity [7–11]. We intend to analyze, in particular, a measure of complexity called Integrated Information - $\Phi$ [6] which has recently gained much popularity under the purview of Integrated Information Theory of Consciousness (IIT) [6]. Though theoretically well founded, IIT 3.0 suffers from several limitations such as current state dependency, computationally expensive and inability to be used with neurophysiological data. There are two other measures viz. neural complexity [12] and causal density [13] as well, which also capture the co-existence of integration and differentiation serving as measures of consciousness [4]. Apart from the individual challenges that these measures have, the common fundamental problem to use them in clinical practise is that they are very difficult to calculate for a network with large number of nodes such as the human brain [4]. In the recent past, a clinically feasible measure of consciousness - Perturbational Complexity Index (PCI) was proposed as an empirical measure of consciousness. PCI has been successfully tested in subjects during wakefulness, dreaming, non-rapid eye movement sleep, anesthesia induced patients, and coma patients. Although the authors of [7] claim that PCI is theoretically based, they don't explicitly and formally establish a link to integration theories.

On one hand we have theoretically well founded measures such as Integrated Information, Causal Density and Neural Complexity, which are currently impossible to be tested in the clinic on a real subject; on the other hand we have the very promising and successful candidate - PCI, which is applicable in the clinic, but lacks a clear connection to these theoretical measures. Our aim is to bridge this gap.

Inspired by the theoretical framework of IIT 3.0 and empirical measure PCI, we propose a compression-complexity measure of integrated information - $\Phi^C$. The idea of *Compression-Complexity* is motivated by observing the similarity between data compression performed by compression algorithms and information integration as performed by the human brain. The link between data compression and Tononi's integrated information is highlighted by the fact that the information encoded by the bits of a compressed file is more than the sum of its parts [14]. Complexity measures based on lossless data compression algorithms such as Lempel-Ziv Complexity (LZ) [15] and Effort-To-Compress (ETC) [16] are known to outperform infotheoretic measures such as entropy for characterizing the complexity of short and noisy time series of chaotic dynamical systems [16]. The newly proposed compression-complexity measure $\Phi^C$ characterizes dynamical complexity (integrated information) of networks using LZ and ETC measures.

$\Phi^C$ is defined and computed as the maximally-aggregate differential normalized Lempel-Ziv (LZ) or normalized Effort-To-Compress (ETC) complexity for the time series data of each node of a network, generated by perturbing each possible atomic bipartition of an $N$-node network with a maximum entropy perturbation and a zero entropy perturbation. $\Phi^C$ has the following advantages - current-state independence, theoretically well-bounded, linearly correlated with entropy of the nodes, and measures integrated information with both aspects - 'process' and 'capacity'. $\Phi^C$ captures the co-existence of differentiation, integration, as well as entropy in networks and shows a similarity with $\Phi$ in its behaviour on 3, 4 and 5-node networks.

# Results

The Results section is categorized as follows: we start by analysing IIT 3.0 and its limitations, in particular, its dependence on current state which makes $\Phi$ a non-robust measure. This limitation is one of the motivations for proposing a new measure. We also demonstrate the correlation between $< \Phi >$ (mean value of $\Phi$) and the entropy of the nodes of the network. In the next section, we allude to the lack of a clear theoretical framework in PCI which makes it an empirical measure. To address these limitations, we first introduce the idea of compression-complexity and then propose a new measure - $\Phi^C$. The steps for the computation of the new measure are provided and its properties are enlisted. We also contrast the hierarchy of $< \Phi >$ with $< \Phi^C >$ for all 3, 4, 5-node networks formed by logic gates: $OR$, $AND$ and $XOR$.

## Model Assumptions

We make the following model assumptions in our paper:

- Although a network can never reach a particular state, we still consider that any state is equally likely at time $t = 0$. Hence, while computing all measures in the paper, we consider all possible current states to be equally likely.

- Each network that we consider is fully connected (bi-directionally) and no node has self-loops unless otherwise specified.
- We assume all networks to be composed of binary logic gates (*OR*, *AND* and *XOR*) and both the perturbation and output time series are also binary. However, our methods can be extended for networks which are non-boolean.
- At certain places in this paper, we have used the term 'element' and 'system' to mean 'node' and 'network' respectively.

## Analysing IIT 3.0 and its limtiations

Integrated Information Theory [6] measures the information that is specified by a system that is irreducible to that specified by its parts. Integrated Information ($\Phi$) is calculated as the distance between the conceptual structure specified by the original system and that specified by its minimum information partition. IIT 3.0 introduces major changes over IIT 2.0 [17–19] and IIT 1.0 [10], but it still suffers certain limitations which shall be discussed.

### Dependence of $\Phi$ on the current state

$\Phi$, as defined in [6], is heavily dependent on the current state of a system. This fact is supported by referring to the framework of IIT 3.0 - (i) firstly, the notion of intrinsic information that Tononi propounds is defined as "difference that make a difference" to a system, which is based on the how an element of a system constrains the past of other node of the same system depending on its mechanism and its current state [6], (ii) secondly, expanding on the notion of integration, the Integrated Information of a mechanism in its current state is computed as the minimum of the past and future integrated information [6], (iii) thirdly, the central identity of IIT 3.0 states that - "an experience is identical with the maximally irreducible conceptual structure (MICS, integrated information structure, or quale) specified by the mechanisms of a complex in a state". Therefore, the conceptual structure is based on the current state of the system [6], (iv) fourthly, the theory goes on to state that certain inactive systems could be conscious as well because consciousness is generated not just by the active elements, but also the inactive elements of a system, (v) lastly, IIT is based on a basic premise that if integrated information has to do something with consciousness, then it must not change, howsoever, the system is divided into its parts. Therefore we require a crucial cut - Minimum Information Partition (MIP) which is the weakest link of the system [20]. This weakest link is dependent on the current state of the system because it requires the identification of the partition which makes least difference to the cause-effect repertoires of the system [6].

Therefore, following from the above, we can infer that $\Phi$ is dependent on the current state of a system. However, this can be problematic as shown in Fig 1. Fig 1(A) shows a system *ABC* with 3 different mechanisms and Fig 1(B) shows different values of $\Phi$ for the different current states of *ABC*, which shows the current state dependence of $\Phi$.

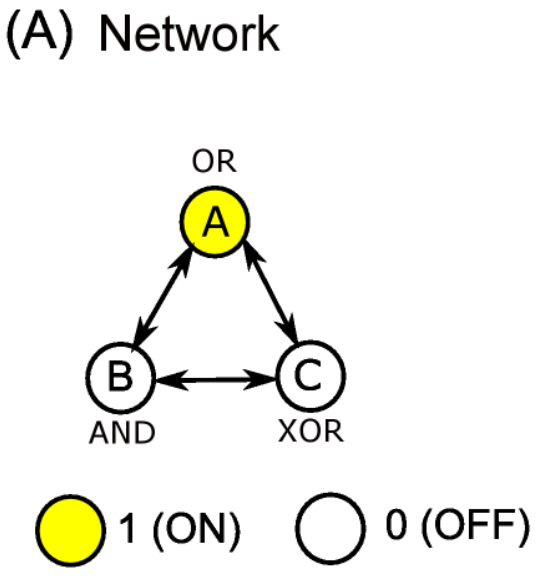

(B) Integrated Information (IIT 3.0)

| Current State | $\Phi$ |
|---|---|
| (0,0,0) | 0.667 |
| (0,0,1) | 0.25 |
| (0,1,0) | 1 |
| (0,1,1) | 1 |
| (1,0,0) | 1.917 |
| (1,0,1) | 1.817 |
| (1,1,0) | 0.25 |
| (1,1,1) | 0.667 |

Fig 1: **Dependence of $\Phi$ on current state.** (A) A 3-node network *ABC* with 3 different mechanisms *OR*, *AND*, *XOR* respectively. (B) The table of values of $\Phi$ for all current states of the network *ABC*.

### $< \Phi >$: Incorporating current states of a network

Taking a cue from the previous section, we performed computer simulations to compute the values of $\Phi$ for all 3-node networks comprising of *OR*, *AND* and *XOR* gates, and for every current state (details in Methods section). We then compute *mean value of* $\Phi$ *across all current states of a network* - $< \Phi >$. Table 1 shows $\Phi$ for all current states, along with the $< \Phi >$ and standard deviation. We repeat this exercise for 4 and 5-node networks as well, and the results are presented in Table 1 of S1 Table.

Table 1: **Integrated Information ($\Phi$) computed for all current states of different 3-node networks.**

| Network No. | Networks | (0,0,0) | (0,0,1) | (0,1,0) | (0,1,1) | (1,0,0) | (1,0,1) | (1,1,0) | (1,1,1) | $< \Phi > \pm$ *Stdev.* |
|---|---|---|---|---|---|---|---|---|---|---|
| 1 | $XOR - XOR - XOR$ | 1.875 | 4.125 | 4.125 | 1.875 | 4.125 | 1.875 | 1.875 | 4.125 | 3 ± 1.203 |
| 2 | $XOR - XOR - OR$ | 2.937 | 3.229 | 4.187 | 0.854 | 4.187 | 0.854 | 2.187 | 2.104 | 2.568 ± 1.312 |
| 3 | $XOR - XOR - AND$ | 2.104 | 2.187 | 0.854 | 4.187 | 0.854 | 4.187 | 3.229 | 2.937 | 2.568 ± 1.312 |
| 4 | $OR - OR - XOR$ | 2.500 | 0.250 | 4.167 | 0.917 | 4.167 | 0.917 | 0.357 | 0.357 | 1.704 ± 1.680 |
| 5 | $AND - AND - XOR$ | 0.357 | 0.357 | 0.917 | 2.042 | 0.917 | 2.042 | 0.250 | 4.500 | 1.422 ± 1.434 |
| 6 | $OR - AND - XOR$ | 0.667 | 0.250 | 1 | 1 | 1.917 | 1.817 | 0.250 | 0.667 | 0.946 ± 0.636 |
| 7 | $AND - AND - OR$ | 0.383 | 0.334 | 0.264 | 0.243 | 0.264 | 0.243 | 0.500 | 0.264 | 0.312 ± 0.091 |
| 8 | $OR - OR - AND$ | 0.264 | 0.500 | 0.243 | 0.264 | 0.243 | 0.264 | 0.334 | 0.383 | 0.312 ± 0.091 |
| 9 | $AND - AND - AND$ | 0.194 | 0.243 | 0.243 | 0.264 | 0.243 | 0.264 | 0.264 | 0.500 | 0.277 ± 0.093 |
| 10 | $OR - OR - OR$ | 0.500 | 0.264 | 0.264 | 0.243 | 0.264 | 0.243 | 0.243 | 0.194 | 0.277 ± 0.093 |

For each possible network formed by three different logic gates: $OR$, $AND$ and $XOR$, the values of $\Phi$ and $< \Phi >$ (± standard deviation) for all 8 current states are calculated. The computation of $\Phi$ is done using Python library for Integrated Information [6,21] which is based on the theoretical framework of IIT 3.0 [22].

$< \Phi >$ exhibits a unique property of integrated information: the hierarchy in its values for all possible 3, 4, 5-node networks formed by all possible combinations of 3 distinct mechanisms: $AND$, $OR$, and $XOR$. As we can observe in Table 1 of S1 Table, $< \Phi >$ leads to a natural hierarchy of networks based on the entropy of its individual nodes and how they combine. The higher the number of high entropy nodes present in the network, the more it contributes to integrated information of the corresponding network (Fig 2). Thus, a 3-node network comprising of all $XOR$s has higher value of $< \Phi >$ (= 3.0) as compared to a network comprising of all $AND$s ($< \Phi >= 0.277$) (please refer S1 Table). It is easy to verify that $XOR$s have the highest Shannon entropy (= 1.0 bit/symbol) followed by $AND$ and $OR$, both of which have an entropy of 0.8113 bits. It is pertinent to note that the natural hierarchy is exhibited by $< \Phi >$ alone and not when the values of $\Phi$ are compared across different networks for any single current state.

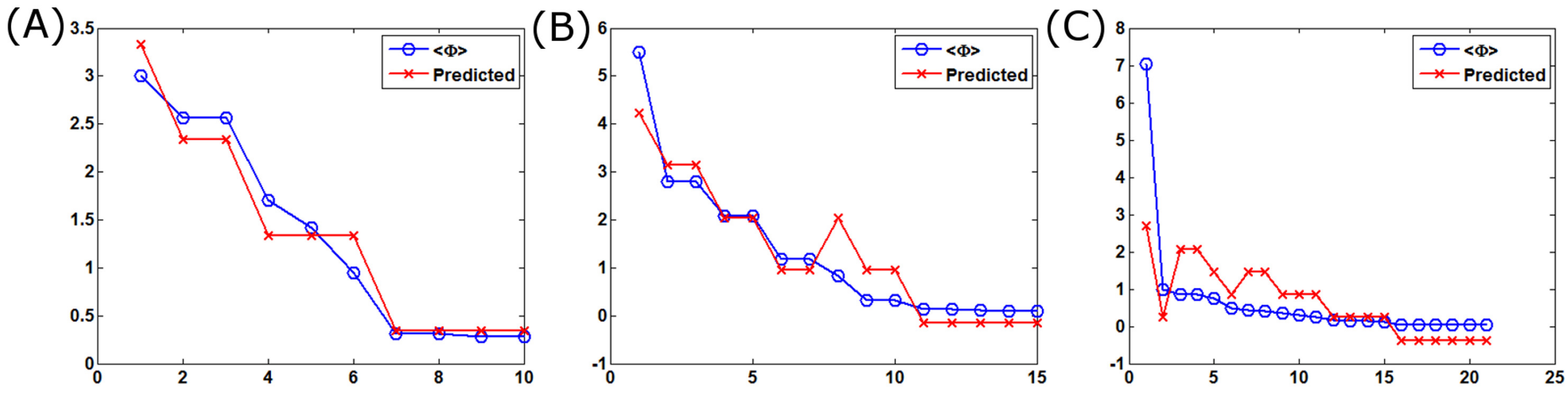

Fig 2: **Linear regression of $< \Phi >$ as a function of entropy of nodes for all 3, 4 and 5-node networks.** A linear fit is obtained between the dependent variable $< \Phi >$ and the explanatory variables 'entropy' of nodes and 'number of nodes'. Y-axis represents the mean value of integrated information in all 3 graphs. (A) X-axis represents 10 different 3-node network configurations (refer Table 1.(a) in S1 Table). (B) X-axis represents 15 different 4-node network configurations (refer Table 1.(b) in S1 Table). (C) X-axis represents 21 different 5-node network configurations (refer Table 1.(c) in S1 Table). The blue plot represents the $< \Phi >$ values for each network configurations and the red plot represents the predicted values of $< \Phi >$ as a function of 'entropy' for each network configuration in all 3 graphs. The predicted $< \Phi >$ obtained from linear regression is a good fit (in red) when compared to the actual $< \Phi >$ (in blue). For further details, please refer to S1 Text.

In order to understand the dependence of $< \Phi >$ with entropy of the nodes, we performed a linear regression (least squares) between the dependent variable $< \Phi >$ and the explanatory variables 'entropy' of the nodes and the 'number of nodes' (for further details, please refer to S1 Text). The predicted values obtained from the linear fit closely tracks the actual values of $< \Phi >$ as shown in Fig 2. This confirms our intuition that there is a linear correlation between the values of $< \Phi >$ and the entropy and of the nodes and their number.

In this section, we have shown that $\Phi$ is heavily dependent on current states of a network, which makes it non-robust measure of integrated information and $< \Phi >$ has linear correlation with the entropy of nodes. $\Phi$ also suffers from the limitations such as computational explosion for estimation in large networks and inability to handle neurophysiological data which is continuous in nature (for ex. time series data) and thus not immediately applicable in the clinic. The new measure $\Phi^C$ that we propose tries to address these limitations.

## Theoretical gap in Perturbational Complexity Index (PCI)

PCI is defined as "the normalized Lempel-Ziv complexity of the spatiotemporal pattern of cortical activation triggered by a direct Transcranial Magnetic Stimulation (TMS) perturbation" [7]. PCI computes the algorithmic complexity of the brain's response to the perturbation and determines two important components of complexity: integration and differentiation, for the overall output of the corticothalamic system. PCI is also different from other measures of complexity for brain signals, in a way that it is resistant to noise from muscle activity or those neuronal sources which don't contribute to integration significantly [7].

Perturbational Complexity Index (PCI) [7] is proposed as an objective clinical measure for the determination of consciousness and for distinguishing the level of consciousness in 3 scenarios: (i) healthy subjects in wakefulness, non-rapid eye movement (NREM) sleep and dreaming states, (ii) subjects who have been induced with sedation by anaesthetic agents (midazolam, xenon, and propofol), and (iii) patients who emerged from coma (vegetative state, minimally conscious state, and locked-in syndrome) [7]. The idea that consciousness originates from complex brain activity patterns which encompasses the fundamental notions of differentiation in space-time (information content) and integration in corticothalamic networks, is considered to be the theoretical basis of PCI [7–11].

PCI faces certain drawbacks which needs to be addressed: a) the authors of PCI have not explicitly shown the mapping between the values of their measure (for example - high in wakefulness and low in NREM sleep) and the amount of integration and differentiation present in the cortical responses, b) PCI measures complexity of averaged TMS evoked potentials from one particular target region (single type of external perturbation) [22], and c) it is not known whether TMS-induced perturbations in PCI are random in nature or not.

Nevertheless, in spite of the individual drawbacks that IIT 3.0 and PCI have, the former is strongly theoretically grounded and latter has succeeded empirically. Inspired by the both of these approaches, we propose new approach based on perturbational compression-complexity, which attempts to bridge the gap between IIT and PCI.

## $\Phi^C$: Moving towards a new approach

To address the above mentioned limitations of $\Phi$ (IIT 3.0) and PCI, we propose a new measure $\Phi^C$ and formally introduce the required steps for its computation. We claim that our proposed measure $\Phi^C$ enables a fast, robust and current-state independent estimation of a measure of integrated information which captures the simultaneous existence of functional differentiation, integration and entropy in networks.

### Data Compression and Integrated Information

As Maguire notes, there is a unique integration of our experience with our existing memories, and this binding gives a subjective flavour to our experience [14]. This fact relates to integrated information. For example, a video camera which is capable of recording several amounts of visual data, is not conscious in the same way as we human beings are [14]. This is because, one can selectively delete the memory of the video camera unit whereas it is nearly impossible to do so in the human brain. The different parts of the brain are tightly integrated such that they have significant causal interactions amongst them and the information of an external stimulus is 'encoded' (or integrated) to the existing information in the brain. Thus, the brain responds more like a singular unified integrated system.

The notion of data compression is a good example for integrated information [23]. In an uncompressed text file, every character is carrying independent information about the text while in a compressed (lossless) file, no single bit is truly independent of the rest. As observed in [14], "*the information encoded by the bits of a compressed* *file is more than the sum of its parts"*, highlighting connections between data compression and Tononi's concept of integrated information.

Compressionism - a term coined by Maguire and Maguire [23, 24], is an attempt to characterize sophisticated data compression carried out by the brain in order to bind information that we associate with consciousness. Therefore, information integration in brain networks could be captured by data compression.

### Compression-Complexity

There is a deep relationship between data compression and several complexity measures, especially those measures which are derived from lossless compression algorithms. Lempel-Ziv complexity (LZ) [15] measures the degree of compressibility of an input string, and is closely related to Lempel-Ziv compression algorithm (a universal compression algorithm [25] which forms the basis of WinZip, Gzip etc.). Similarly, a recently proposed complexity measure known as Effort-To-Compress (ETC) [16] characterizes the effort to compress an input sequence by using a lossless compression algorithm. The specific compression algorithm used by ETC is Non-Sequential Recursive Pair Substitution Algorithm (NSRPS) [26]. ETC and LZ have been demonstrated to outperform Shannon entropy for characterizing the complexity of short and noisy time series from chaotic dynamical systems [16]. It is difficult to evaluate entropy since it involves estimation of probability distribution which requires extensive sampling that usually cannot be performed [27]. However, LZ and ETC complexities are properties of individual sequences (or time series) and much easier to compute in a robust fashion.

In the light of the above advantages which LZ and ETC provide over information theoretic measures such as entropy, we are motivated to employ these in characterizing integrated information. Therefore, we introduce "*Compression-Complexity"* measures which characterize dynamical complexity of brain networks using lossless compression algorithm based complexity measures.

Our goal is to use these complexity measures (LZ and ETC) to quantify the amount of integrated information in a network. When a single node of a network is perturbed by a random input, this perturbation travels through the network to other nodes. By capturing the output at all the other nodes and computing the complexity of their outputs, we intend to study the degree of information integration in the network. As a baseline, we also compute the complexity of the response of all the other nodes for a zero-entropy perturbation of the input node. We then compute the difference between the two responses and aggregate them. A network which is more strongly integrated will exhibit strong causal interactions among its nodes. This means that in such a network, the perturbations travel throughout the network causing high entropy output in other nodes as well (since the input is a random perturbation, it is a high entropy input to the network). By *aggregating* the *differential* compression-complexity of the output of all the other nodes (leaving out the input node which is perturbed), we get a sense of integrated information. This is because, we are computing information as *difference that makes a difference* [28], here the difference is calculated between the response for a maximum-entropy perturbation and a zero-entropy perturbation. We then take a maximum of all such aggregated differential compression-complexity measures across all possible perturbations (if a network has $N$ nodes, then we have $N$ pairs of perturbations in total). The reason for taking the maximum is that it indicates that specific atomic bipartition which characterizes integrated information as *maximum difference in the input perturbations (in terms of entropy) that makes a maximum difference in the aggregated output response (in terms of compression-complexity)*. Thus, we define *the maximum differential compression-complexity (aggregated) response that is triggered by a maximum differential entropy perturbation* as a measure of the capacity of the network to integrate information. In a way, this is what PCI is also measuring, but it makes use of a single perturbation (which is not a maximum entropy one either).

### Defining and Computing the new measure $\Phi^C$

$\Phi^C$ for a network (with randomly chosen current state of the network) is computed by performing the following steps, as also depicted in Fig 3: (i) bipartitioning a network into its all atomic bipartitions, (ii) perturbing the atomic node for each bipartition with random input time series (maximum entropy), and followed by a zero entropy time series (constant sequence), (iii) recording the output time series from all the other nodes of the network and computing the complexity of these individual time series using LZ/ETC for each bipartition, for both perturbations and computing their difference (denoted by ${}_{\text{LZ}}\varphi^{\text{C}}$ or ${}_{\text{ETC}}\varphi^{\text{C}}$), (iv) computing the aggregate of these differential complexity measures (LZ/ETC) for each bipartition of network, (v) reporting the maximum value out of all such computed aggregate differential complexity measures (${}_{\text{LZ}}\phi^{\text{C}}$ or ${}_{\text{ETC}}\phi^{\text{C}}$) obtained in step (iv) as the value of ${}_{\text{LZ}}\Phi^{\text{C}}$ (or ${}_{\text{ETC}}\Phi^{\text{C}}$).

**Definition:** $\Phi^C$ *is defined as the maximally-aggregate differential normalized Lempel-Ziv (LZ) or normalized Effort-To-Compress (ETC) complexity for the time series data of each node of a network, generated by perturbing* *each possible atomic bipartition of an $N$-node network with a maximum entropy perturbation and a zero entropy perturbation.* The mean of $\Phi^C$ across all states of a network is denoted as $< \Phi^C >$. ${}_{ETC}\Phi^C$ and ${}_{LZ}\Phi^C$ denote $\Phi^C$ computed using ETC and LZ complexity measures respectively.

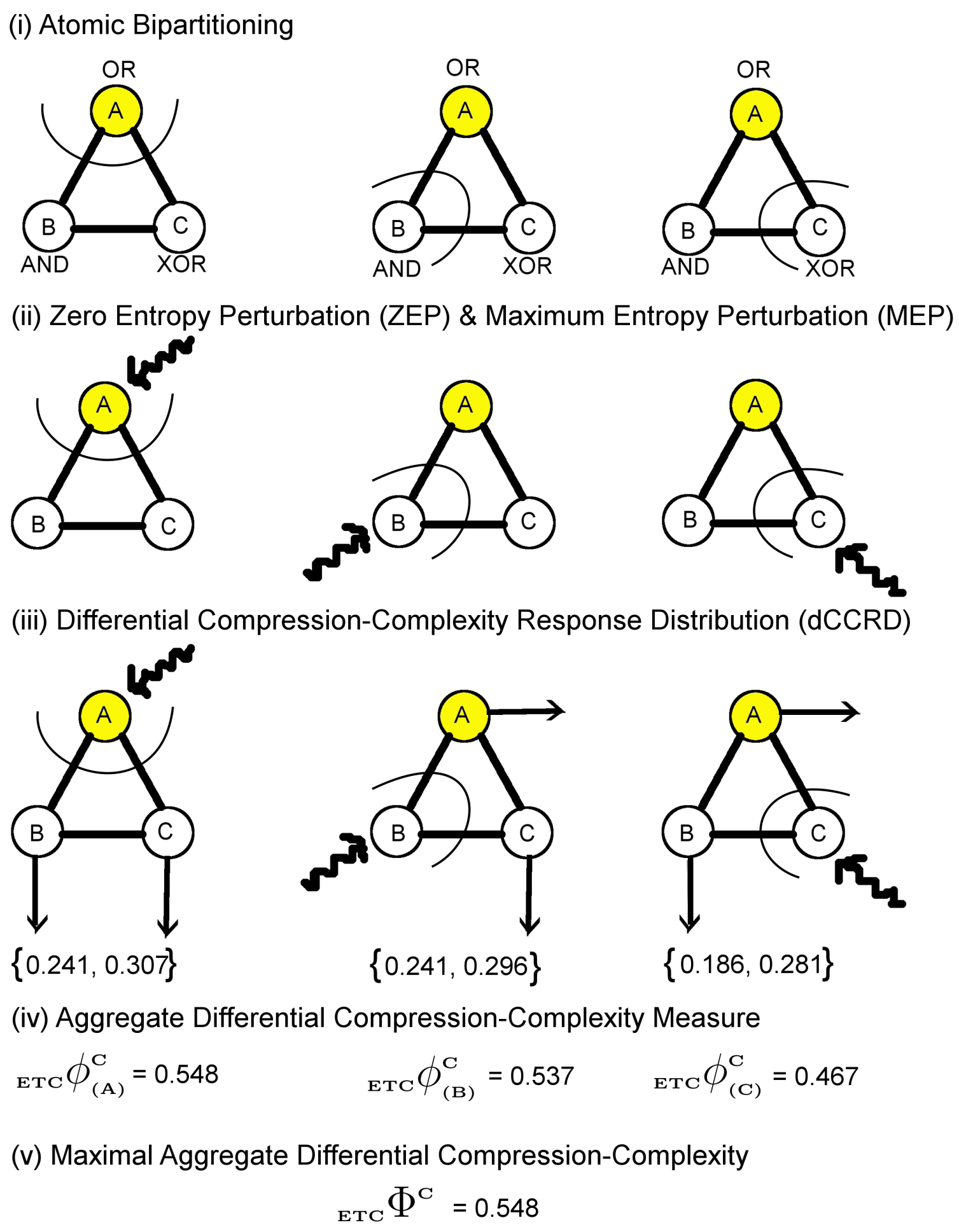

Fig 3: **Algorithm for the computation of $\Phi^C$ is illustrated through diagrams.** The network $ABC$ (current state $= (1,0,0)$) constitutes three logic gates: $OR$, $AND$, $XOR$ for which the value of $\Phi^C$ is computed. (i) The network is partitioned into 3 possible atomic bipartitions, (ii) each atomic bipartition is perturbed with a Maximum Entropy Perturbation (MEP) which is a random input binary time series (length$=200$) as well as Zero Entropy Perturbation (ZEP) which is a constant sequence (length$=200$), (iii) Differential Compression-Complexity is computed by taking the difference between complexities for MEP and ZEP for each output time series from the remaining two unperturbed nodes. This forms the Differential Compression-Complexity Response Distribution (dCCRD) for each bipartition. For example, $\{{}_{\mathrm{ETC}}\varphi^{\mathrm{C}}_{\mathrm{B(A)}} = 0.241,\ {}_{\mathrm{ETC}}\varphi^{\mathrm{C}}_{\mathrm{C(A)}} = 0.307\}$, represents the dCCRD of the time series obtained from the nodes $B$ and $C$ respectively, when the node $A$ is perturbed. Similarly, the dCCRD for the other two bipartitions are: $\{{}_{\mathrm{ETC}}\varphi^{\mathrm{C}}_{\mathrm{A(B)}} = 0.241,\ {}_{\mathrm{ETC}}\varphi^{\mathrm{C}}_{\mathrm{C(B)}} = 0.296\}$ and $\{{}_{\mathrm{ETC}}\varphi^{\mathrm{C}}_{\mathrm{A(C)}} = 0.186,\ {}_{\mathrm{ETC}}\varphi^{\mathrm{C}}_{\mathrm{B(C)}} = 0.281\}$, (iv) the individual values of each dCCRD are summed up to obtain 'Aggregate Differential Compression-Complexity Measure' for each bipartitioned-perturbed network. Therefore, ${}_{\mathrm{ETC}}\phi^{\mathrm{C}}_{(\mathrm{A})} = {}_{\mathrm{ETC}}\ \varphi^{\mathrm{C}}_{\mathrm{B(A)}} + {}_{\mathrm{ETC}}\ \varphi^{\mathrm{C}}_{\mathrm{C(A)}}$ and similarly ${}_{\mathrm{ETC}}\phi^{\mathrm{C}}_{(\mathrm{B})}$ and ${}_{\mathrm{ETC}}\phi^{\mathrm{C}}_{(\mathrm{C})}$ can be computed. All corresponding values are: ${}_{\mathbf{ETC}}\phi^{\mathbf{C}}_{(\mathbf{A})} = 0.548$, ${}_{\mathbf{ETC}}\phi^{\mathbf{C}}_{(\mathbf{B})} = 0.537$, ${}_{\mathbf{ETC}}\phi^{\mathbf{C}}_{(\mathbf{C})} = 0.467$, (v) Maximal-Aggregate Differential Compression-Complexity, $\Phi^C$, is nothing but the maximum of the Aggregate Differential Compression-Complexity measures: $\max({}_{\mathrm{ETC}}\phi^{\mathrm{C}}_{(\mathrm{A})}, {}_{\mathrm{ETC}}\ \phi^{\mathrm{C}}_{(\mathrm{B})}, {}_{\mathrm{ETC}}\ \phi^{\mathrm{C}}_{(\mathrm{C})})$. Thus, ${}_{\mathbf{ETC}}\Phi^{\mathbf{C}} = 0.548$. For more details, refer S1 Text.

For the sake of clarity and completeness, we define the following terms:
**Network**: A system with $N$ nodes $A_1, A_2, \ldots, A_N$ with all bi-directional connections and no self-loops.

**Atomic bipartition**: A division of a network with two parts with one part containing only one node ($A_i$) and the other part containing the rest $\{A_1, A_2, \dots, A_j, \dots, A_N\}$ where $j \neq i$.
**Maximum Entropy Perturbation (MEP)**: It is defined as the uniform random input perturbation time series (with maximum entropy) injected to $A_i$ of the atomic bipartition.
**Zero Entropy Perturbation (ZEP)**: It is defined as a constant input perturbation time series (with zero entropy) injected to $A_i$ of the atomic bipartition.
**Differential Compression-Complexity Response Distribution (dCCRD)**: It is defined as the distribution of difference between complexities of the responses from each node of the network in each atomic bipartition of the network when one of the nodes is perturbed - first with a random maximum entropy perturbation and next with a zero entropy perturbation (see Methods for details).

### An example of $\Phi^C$

$\Phi^C$ serves as a measure of integrated information (similar to $\Phi$). We provide two examples to demonstrate the correspondence of $\Phi^C$ with $\Phi$. For two 2-node networks as shown in Fig 4, the values of $\Phi^C$ and $\Phi$ are similar - both are lower for $OR - AND$ than $OR - XOR$ network.

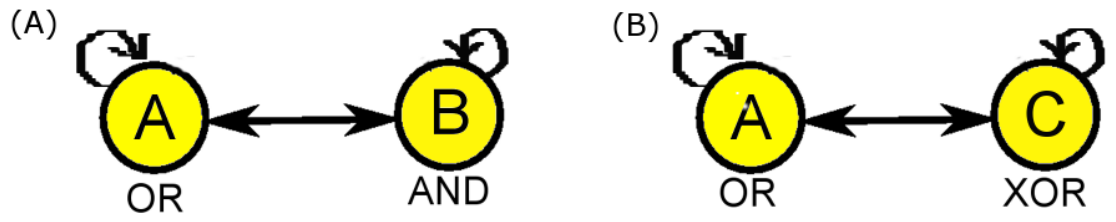

(C)

| Networks | $\Phi$ | $_{ETC}\Phi^C$ | $_{LZ}\Phi^C$ |
|---|---|---|---|
| $OR - AND$ | 0.090 | 0.033 | 0.019 |
| $OR - XOR$ | 0.188 | 0.302 | 1.103 |

Fig 4: **Resemblance of $\Phi^C$ with $\Phi$ for two 2-node networks.** (A): $OR - AND$ network, (B) $OR - XOR$ network. (C) The table lists the corresponding values of $\Phi^C$ and $\Phi$ for the current state $(1, 1)$. It can be seen that similar to $\Phi$, $\Phi^C$ is lower for $OR - AND$ when compared to $OR - XOR$.

## Comparing $< \Phi^C >$ with $< \Phi >$

In this section, we intend to evaluate how $< \Phi^C >$ does in comparison with $< \Phi >$ for 3, 4, 5-node networks. It is shown through simulations that $< \Phi^C >$ aligns very well with $< \Phi >$ in terms of hierarchy for 3-node networks and to a certain extent with 4 and 5-node networks as shown in S1 Table and Fig 5.

The trends in the values of $< \Phi >$ and $< \Phi^C >$ across different networks is depicted in Fig 5 and they are similar. Also, as shown in Fig 6, we depict box-plots of the values of $\Phi$ and $\Phi^C$ for all networks and for all current states.

For the sake of exhaustive analysis, we present mean and standard deviation of $\Phi^C$ and $\Phi$ for all current-states of each network ( S1 Table). $< \Phi^C >$ is observed to have similar hierarchy as $< \Phi >$ but with lesser standard deviation across current-states for all 3, 4, 5-node networks. As depicted in Tables 1.(a), 2.(a) and 3.(a) in S1 Table, 3-node networks exhibit a similar hierarchy in values of $<_{ETC} \Phi^C >$ and $<_{LZ} \Phi^C >$ when compared to the values of $< \Phi >$. This order is found to some extent in 4 and 5-node networks (refer to Tables in S1 Table). However, there are some departures in the ordering of $< \Phi^C >$ and $< \Phi >$. For example, while comparing $< \Phi >$ and $<_{LZ} \Phi^C >$ and taking the $< \Phi >$ values in Table 1 (S1 Table) as a reference for 3-node networks, there is a minor shuffling in the hierarchy (this is clear when you look at the column 'Network No.'). For 4-node and 5-node, the departure from the hierarchy with respect to $< \Phi >$ is higher. As an example, $AND - AND - AND - AND - AND$ and $AND - AND - AND - AND - XOR$ stand at #20 and #2 respectively in the hierarchy for $< \Phi >$, whereas for $< \Phi^C >$ they are much closer in hierarchy. It is more intuitive that a single $XOR$ replacement of an $AND$ gate should not yield such a drastic change in integrated information.

Also, the standard deviations of $\Phi^C$ for 3, 4 and 5-node networks is much lower than that of $\Phi$: $(0.0 - 0.184)$ for $_{ETC}\Phi^C$, $(0.0 - 0.558)$ for $_{LZ}\Phi^C$ and $(0.0 - 2.062)$ for $\Phi$. In order to measure the dispersion of the three measures across all networks and all states, we compute the coefficient of variation (CoV) defined as the ratio of standard deviation to the mean. This is plotted in Fig 7, from which it is evident that both $_{LZ}\Phi^C$ and $_{ETC}\Phi^C$ have better (lower) values of CoV than $\Phi$, barring a few exceptions (network #9, 10 for 3-node networks, #14, 15 for 4-node networks and #14, 18, 20, 21 for 5-node networks). Therefore, in practice, we recommend choosing any single current state at random and then computing the value of $\Phi^C$ for that current state. This is also one of the reasons why our measure is computationally efficient.

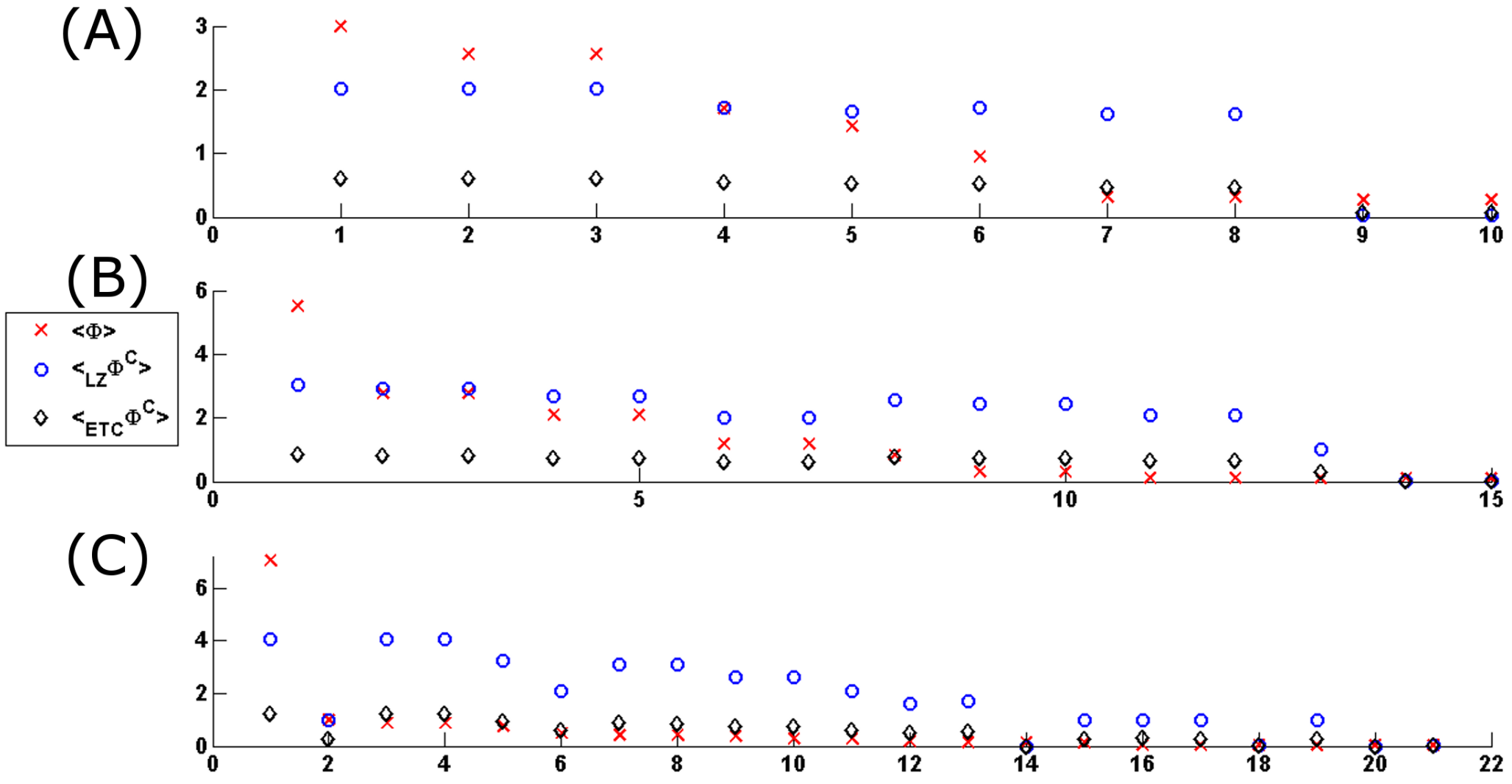

Fig 5: **Plots of $<\Phi^C>$ and $<\Phi>$ (across all current-states) for all (A) 3-node, (B) 4-node, (C) 5-node networks.** X-axis of each graph represents the different configurations of networks and Y-axis represents mean values of integrated information corresponding to the tables in S1 Table. (A), (B) and (C) show the mean value of integrated information for ten configurations 3-node networks, 15 configurations of 4-node networks, 21 configurations of 5-node networks respectively. The trends in the values of $<\Phi>$ and $<\Phi^C>$ across different networks, depicted in this figure, are similar.

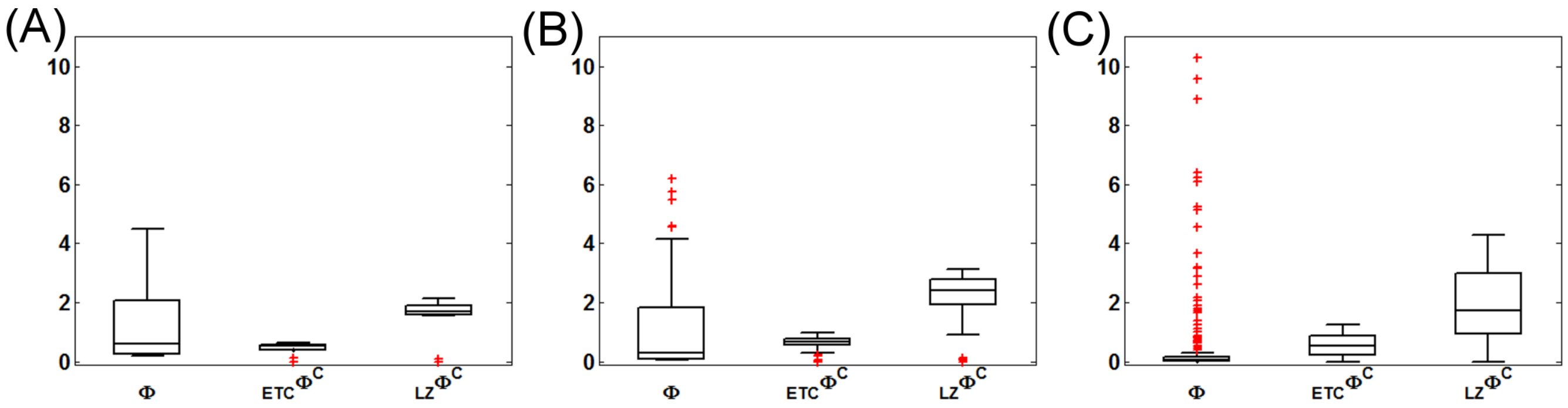

Fig 6: **Box-plots of the values $\Phi$, ${}_{ETC}\Phi^C$ and ${}_{LZ}\Phi^C$ for all (A) 3, (B) 4, (C) 5-node networks and for all current states.** The resolution of $<\Phi>$ across different networks is best among all the three measures.

## Properties of $\Phi^C$

1. **Current-state Independence:** Unlike other measures of integrated information such as causal density [13], neural complexity [12], $\Phi$ (IIT 1.0) [10], $\varphi$ (IIT 2.0) [17,18,29], $\Phi^{Max}$ (IIT 3.0) [5,6], $\Phi^*$ and $\Phi^*_{MMP}$, which demonstrate the state-dependence of integrated information, the proposed measure $\Phi^C$ has negligible dependence on the current state of the nodes of the network. There have been earlier attempts to propose a state-independent measure: (i) $\Phi_E/\Phi_{AR}$ proposed by [29] aims to measure the average information generated by the past states rather than information produced by the particular current state, (ii) $\psi$ proposed by Griffith [30] also suggests stateless $\psi$ as $<\psi>$, but this results in weakening of $\psi$, (iii) $\Phi^{AR}_{MMP}$ suggested by Toker et al. [31] based on the foundations of $\Phi_{AR}$ using Maximum Modularity Partition seems to be state-independent when utilized for neural data that cannot be transformed into a normal distribution. But, these measures too, have not been extensively tested with different networks to show a lower standard deviation when computed across all current states. However, as it can be seen from S1 Table, the standard deviation of the values of $\Phi^C$ across all current states for 3, 4, 5-node networks is very low. We expect this property to hold even for networks with larger number of nodes.

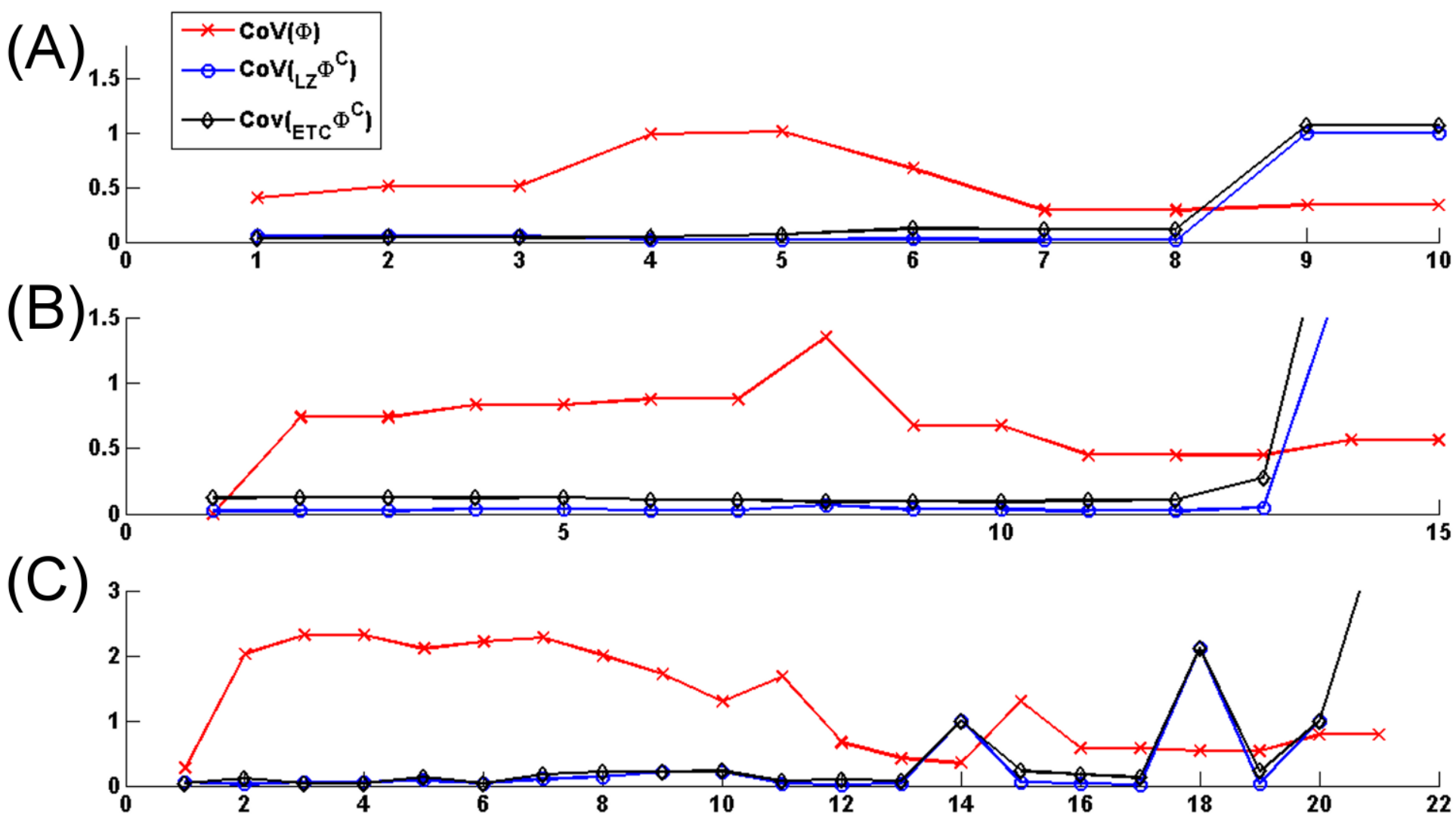

Fig 7: **Coefficient of variation (CoV) for integrated information measures.** CoV of ${}_{LZ}\Phi^C$, ${}_{ETC}\Phi^C$ and $\Phi$ for (A) 3-node, (B) 4-node, and (C) 5-node networks. X-axis of each graph represents the different configurations of networks and Y-axis represents CoV values. (A), (B) and (C) show the mean value of integrated information for ten configurations of 3-node networks, 15 configurations of 4-node networks, 21 configurations of 5-node networks respectively (refer S1 Table for the network configurations). Both ${}_{LZ}\Phi^C$ and ${}_{ETC}\Phi^C$ have better (lower) values of CoV than $\Phi$ in most networks barring a few exceptions.

2. **Linear correlation of $\Phi^C$ with entropy of nodes:** Similar to $< \Phi >$, $< \Phi^C >$ also exhibits a linear correlation with the entropy of the nodes. As shown in Fig 8, linear regression (least squares) is performed with the dependent variable $< \Phi^C >$ and the explanatory variables 'entropy' of the nodes and the 'number of nodes' (for further details, please refer to S1 Text). The predicted values obtained from the linear fit closely tracks the actual values of $< \Phi^C >$ as shown in Fig 8. In fact, the prediction improves as the number of nodes increases.

3. **Information Theoretic vs. Compression-Complexity Measure:** Existing measures of integrated information are all heavily based on information theoretic measures such as entropy, mutual information, intrinsic information etc. However, $\Phi^C$ is built on complexity measures ($ETC$, $LZ$) which have roots in lossless compression algorithms. $ETC$ is related to a lossless compression scheme known as NSRPS [16, 26] and $LZ$ is based on a universal compression algorithm [25]. These complexity measures do not directly model the probability distribution of potential past and future states of a system, but learn from the patterns in the time series. This approach is known to be more robust even with small set of measurements and in the presence of noise [16].

4. **Boundedness:** $\Phi^C$ is well defined mathematically and is bounded between 0 and $N-1$, where $N$ is the number of nodes in the network. Since we use normalized values for both $ETC$ and $LZ$ complexity measures to define $\varphi^C$ at every node, therefore $\varphi^C$ is bounded between 0 and 1. Further, since $\Phi^C$ is computed as the maximum of aggregated values of $\varphi^C$, and for every atomic bipartition there are $N-1$ pairs of output time series, the maximum aggregated value of the differential complexity measure can be utmost $N-1$ (the maximum value is attained if complexity value obtained from MEP time series is 1 and 0 from ZEP time series for each bipartition). Therefore, $0 \leq \Phi^C \leq N-1$. Even though $LZ$ complexity is also normalized, its value can exceed one at times [32, 33]. This is a problem due to finite data lengths. But, normalized $ETC$ does not have this problem and it is always bounded between 0 and 1 [16].

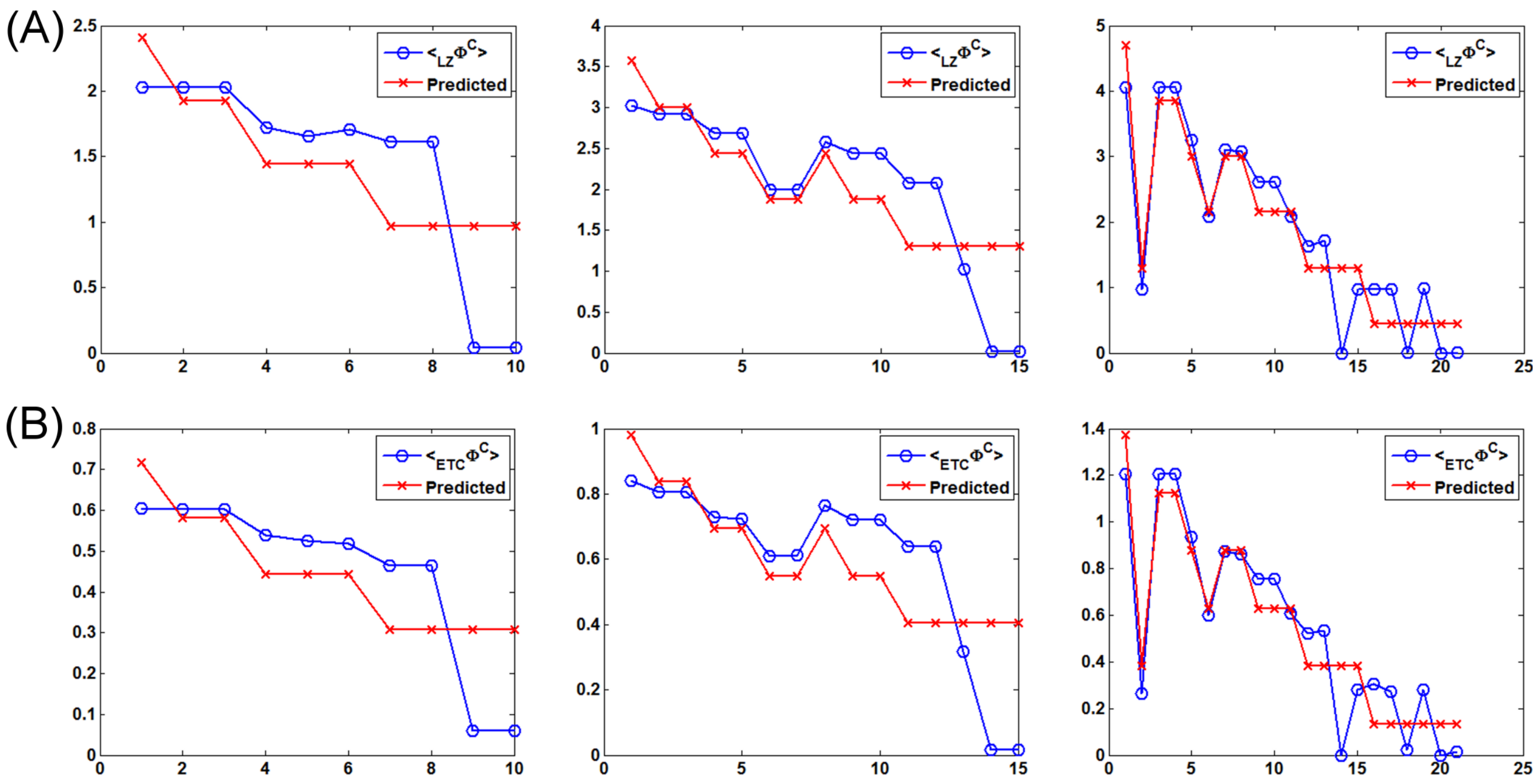

Fig 8: **Linear regression of (A) $<_{LZ}\ \Phi^C>$ and (B) $<_{ETC}\ \Phi^C>$ as a function of entropy of nodes for all 3, 4 and 5-node networks.** A linear fit is obtained between the dependent variable $<_{LZ}\ \Phi^C>$ (or $<_{ETC}\ \Phi^C>$) and the explanatory variables - 'entropy' of nodes and 'number of nodes'. In each of the graphs above, X-axis of each graph represents the different configurations of networks and Y-axis represents the mean value of integrated information. The leftmost, middle and rightmost graph in both (A) and (B) shows the mean value of integrated information for ten configurations 3-node networks, 15 configurations of 4-node networks, 21 configurations of 5-node networks respectively (refer S1 Table for the network configurations). For each network configuration, the blue plot represents $<_{LZ}\ \Phi^C>$ or $<_{ETC}\ \Phi^C>$ values respectively in (A) and (B) and the red plot represents their predicted values as a function of 'entropy'. For further details, please refer to S1 Text.

5. **Process vs. Capacity:** $\Phi^{Max}$ measures consciousness as integrated information which is represented by the capacity of the system [29], while PCI measures the same as a process by recording the activity of the brain generated by perturbing the cortex with TMS using high-density electroencephalography [7]. However, $\Phi^C$ as a measure of integrated information encapsulates both the ideas of 'capacity' and 'process'. The Differential Compression-Complexity Response Distribution (dCCRD) for each atomic bipartition is measuring integrated information as a process for time-series data from each node. The Aggregate Differential Compression-Complexity Measure captures the network's capacity to integrate information. Therefore, $\Phi^C$ serves as a connection between IIT and PCI based approaches of measuring consciousness.

6. **Discrete and Continuous Systems:** $\Phi^C$ can be easily extended to continuous measurements such as neurophysiological data. We could sample the continuous measurements to yield discrete samples on which $\Phi^C$ can be estimated. Thus, our measure applies equally to both discrete and continuous systems.

# Discussion

In this paper, we proposed a new measure for quantifying integrated information (a potential measure of consciousness) called $\Phi^C$, which is defined as the largest aggregated differential compression-complexity measure (ETC/LZ) computed from time series data of each perturbed node of the atomic bipartition of an $N$-node network. We have discussed the motivation behind such a compression-complexity approach to measure integrated information. The perturbational perspective to measure compression-complexity is inspired by PCI and is also computationally ef-ficient (we need to consider only $N$ bipartitioned perturbations). $\Phi^C$ is a measure of the *maximum difference in complexity of outputs resulting from a maximum difference in entropy of input perturbations* across all nodes of a network. $\Phi^C$ exhibits the following salient innovations: (i) negligible current state dependence (as indicated by a very low standard-deviation of $\Phi^C$ across all current states of a network), (ii) integrated information measured as compression-complexity rather than as an infotheoretic quantity, and (iii) quick computation by a perturbational approach over atomic bipartitions (which scales linearly with number of nodes), thus avoiding combinatorial explosion. Our computer simulations showed that $< \Phi^C >$ has similar hierarchy to $< \Phi >$ for 3, 4, 5-node networks, thus conforming with IIT. Moreover, the hierarchy of $< \Phi^C >$ follows intuitively from our understanding that integrated information is higher in a network which has more number of high entropy nodes (for ex. more number of $XOR$ gates than $AND$, $OR$ gates) for a fully connected network.

## Advantages of $\Phi^C$

Our novel approach provides several advantages over other measures of integrated information: i) suggesting atomic bipartitioning instead of MIP which avoids combinatorial explosion, ii) introducing Maximum Entropy Perturbation (MEP) and Zero Entropy Perturbation (ZEP), and iii) proposing Differential Compression-Complexity Response Distribution (dCCRD) allowing us to measure $\Phi^C$ for continuous time series data.

$\Phi^{Max}$ as a measure of Integrated Information to quantify consciousness needs the identification of Minimum Information Partition (MIP) in a network [6]. But, finding MIP faces practical and theoretical roadblocks which are unresolved till now [22]. The practical issue is: locating MIP requires investigation of every possible partition of the network, which is realistically unfeasible as the total number of possible partitions increase exponentially with the size of the network leading to combinatorial explosion [22, 31, 34]. In fact, this approach is impractical for a network with more than a dozen nodes [6]. In order to overcome these issues, other approaches have been suggested, such as Minimum Information Bipartition (MIB) and Maximum Modularity Partition (MMP). Though MIB is faster to compute than MIP [31] and has been used by various measures of integrated information [8, 10, 18, 22, 35–37], it also has two issues to be addressed. Firstly, the time to find MIB also grows exponentially with larger networks and secondly, it is not certain if MIB is a reasonable approach to disintegrate a neural network (since it is dubious that functional subnetworks divide the brain exactly in half.) [31]. Hence, MIB is inapplicable to real brain networks as of now. We tackle this practical issue by using atomic bipartitions, whose number increases linearly with the size of the network. Atomic bipartitions have been recommended by other researchers too in lieu of MIP [22, 31].

Compression-Complexity approach conferred certain desirable properties to $\Phi^C$. Firstly, this approach allowed us to measure the integrated information as a process for the output time-series data in the form of distribution of differential responses (dCCRD) to Maximum Entropy Perturbation (MEP) and Zero Entropy Perturbation (ZEP) and secondly, dCCRD provided us with the distribution of differential complexity values which could be useful in multitude of ways to be explored in the future. Furthermore, since $\Phi^C$ employs complexity measures such as LZ and ETC instead of infotheoretic quantities (such as entropy, mutual information etc.), it is more robust to noise, and efficient with even short and non-stationary measurement time series. Also, we have already noted that $\Phi^C$ has negligible dependence on current-state of a network, unlike other measures.
Thus, $\Phi^C$ is a potentially promising approach for fast and robust empirical computation of integrated information.

## Interplay between differentiation, integration and entropy

Researchers have already acknowledged that consciousness could be a result of the complexity of neuronal network in our brain which depicts 'functional differentiation' and 'functional integration' [7–11, 29, 38]. For example, referring to Fig 9, when we compare the two networks (i) and (iii) with the network (ii), we note that the latter is more heterogeneous (since it has three different types of gates as opposed to the former which has only two types of gates). Griffith [30] makes the point that in such a scenario, it is intuitive that the integrated information is larger for the more heterogeneous network. But, it is not as intuitive as it seems, since the entropy of the gates play an important role as well.

As shown in the Fig 9, the integrated information ($< \Phi >$, $<_{LZ} \Phi^C >$ and $<_{ETC} \Phi^C >$) of the network $AABB$ is lower than that of $AABC$ which is in turn lesser than the integrated information of $AACC$ (with $A = OR$, $B = AND$, $C = XOR$). This may appear counter-intuitive at first, but it makes sense when we realize that the entropy of $C$ is higher than both $A$ and $B$. Thus, it is not universally true that heterogeneous networks have higher amounts of integrated information, as it very much depends on the entropy of the individual nodes as well as their number. In the case of the brain, cortical neurons are known to exhibit different firing patterns whose entropy varies widely. As an example, we simulate a cortical neuron from the Hindmarsh-Rose neuron model [39]

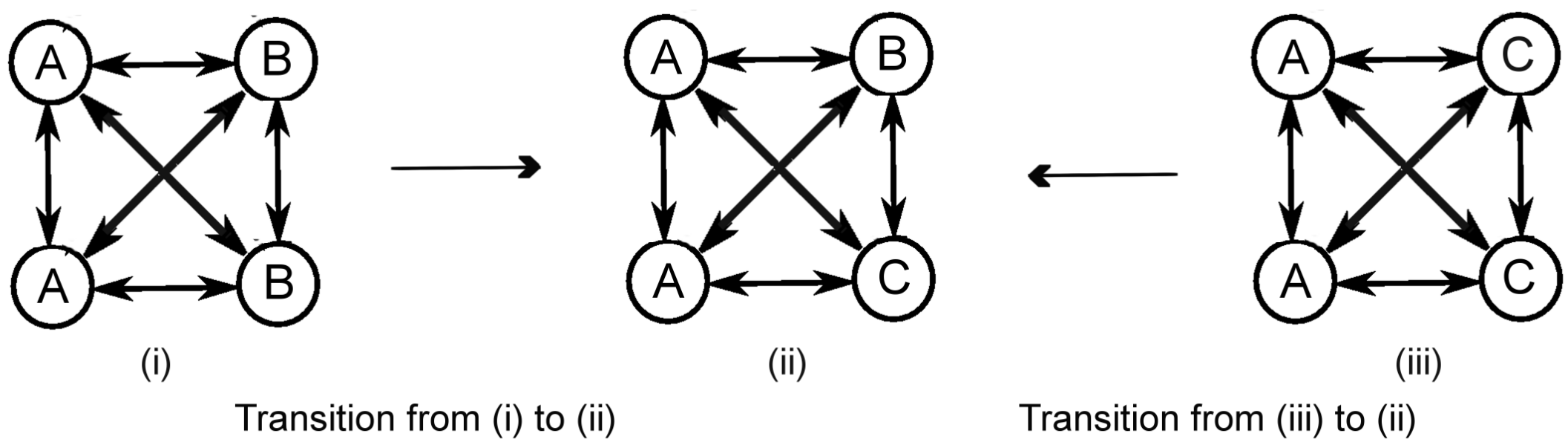

| Networks | $< \Phi >$ | $<_{ETC} \Phi^C >$ | $<_{LZ} \Phi^C >$ |
|---|---|---|---|
| AABB | 0.119 | 0.318 | 1.020 |
| AABC | 0.325 | 0.720 | 2.444 |

| Networks | $< \Phi >$ | $<_{ETC} \Phi^C >$ | $<_{LZ} \Phi^C >$ |
|---|---|---|---|
| AACC | 2.083 | 0.730 | 2.683 |
| AABC | 0.325 | 0.720 | 2.444 |

Note: A = OR, B = AND, C = XOR

Fig 9: **Interplay between differentiation, integration and entropy.**
(i) $AABB$ has $< \Phi >= 0.119$, $<_{ETC} \Phi^C >= 0.318$, $<_{LZ} \Phi^C >= 1.020$, (ii) $AABC$ has $< \Phi >= 0.325$, $<_{ETC} \Phi^C >= 0.720$, $<_{LZ} \Phi^C >= 2.444$, (iii) $AACC$ has $< \Phi >= 2.083$, $<_{LZ} \Phi^C >= 0.730$, $<_{ETC} \Phi^C >= 2.683$. The integrated information of the network $AABB$ is lower than that of $AABC$, which is lesser than the integrated information of $AACC$. This may seem counter-intuitive, but it is not, since the entropy of $C$ ($XOR$ gate) is higher than the entropies of both $B$ ($AND$ gate) and $A$ ($OR$) gate. Thus, heterogeneity alone is insufficient to increase the value of integrated information of the network; the entropy of the individual nodes and their number in the network also matter.

which is a widely used model for bursting-spiking dynamics of the membrane voltage of a single neuron (refer S1 Text). The same neuron exhibits regular spiking (Fig 10(A)) when the external current applied is $I = 3.31$ and chaotic or irregular spiking (Fig 10(B)) when $I = 3.28$. We computed the Shannon entropy, $ETC$, and $LZ$ complexity values for the two cases. It can be seen that the same neuron shows a lower value of entropy and complexities ($H = 0.8342$ bits, $ETC = 0.1910$ and $LZ = 0.6879$) when it is spiking in a regular manner as compared to its behavior in a chaotic manner ($H = 0.9295$ bits, $ETC = 0.2211$, $LZ = 0.7262$). Thus, for the same neuronal network, under two different excitations, the neurons can behave with different entropies/complexities. This will have a significant impact on the values of integrated information and it is hard to predict how this interplay between functional integration, differentiation and entropy will pan out in reality.

## Limitations and Future Work

Though $\Phi^C$ provides certain benefits over other measures of integrated information, it has some shortcomings as well.

1. For 3 and 4-node networks, $< \Phi^C >$ values ($<_{LZ} \Phi^C >$ and $<_{ETC} \Phi^C >$) show poor resolution compared to $< \Phi >$ across various networks as depicted in Fig 6. The reason for this may be the fact that we have considered fully connected networks and the perturbations travel to all parts of the network. Further experiments with different kinds of networks are needed to make conclusive inferences.

2. Even though number of required perturbations for atomic bipartitions scale linearly with the increase in the number of nodes, it is still a mammoth task to perturb all atomic bipartitions for a larger network like the human brain. It is important to note that PCI could still differentiate between different levels of consciousness in wakefulness, sleep, anaesthesia-induced patients etc. though *"it measures the complexity of averaged neural responses to one particular type of external perturbation (e.g. a TMS pulse to a target region)"* [22], rather than all possible perturbations. A heuristic approach to determine the right number of bipartitions and perturbations for evaluating $\Phi^C$ would be a trade-off between our current approach and PCI.

$\Phi^C$ demonstrated various salient innovations and properties which positions it uniquely among the medley of other measures of integrated information (Table 2). But, following are the areas in which future work is required: (i) we did not discuss the relationship between quality of consciousness (phenomenal properties of experience) and properties of $\Phi^C$, (ii) determining $\Phi^C$ for networks with varied connectivity matrices and topologies to understand its behaviour as the configuration of the network changes or size of the network increases, (iii) using $\Phi^C$ on real neural recordings from the brain, (iv) determining an optimal bipartition for computing $\Phi^C$ and then comparing the results, and (v) investigating the application of $\Phi^C$ to networks from other domains.

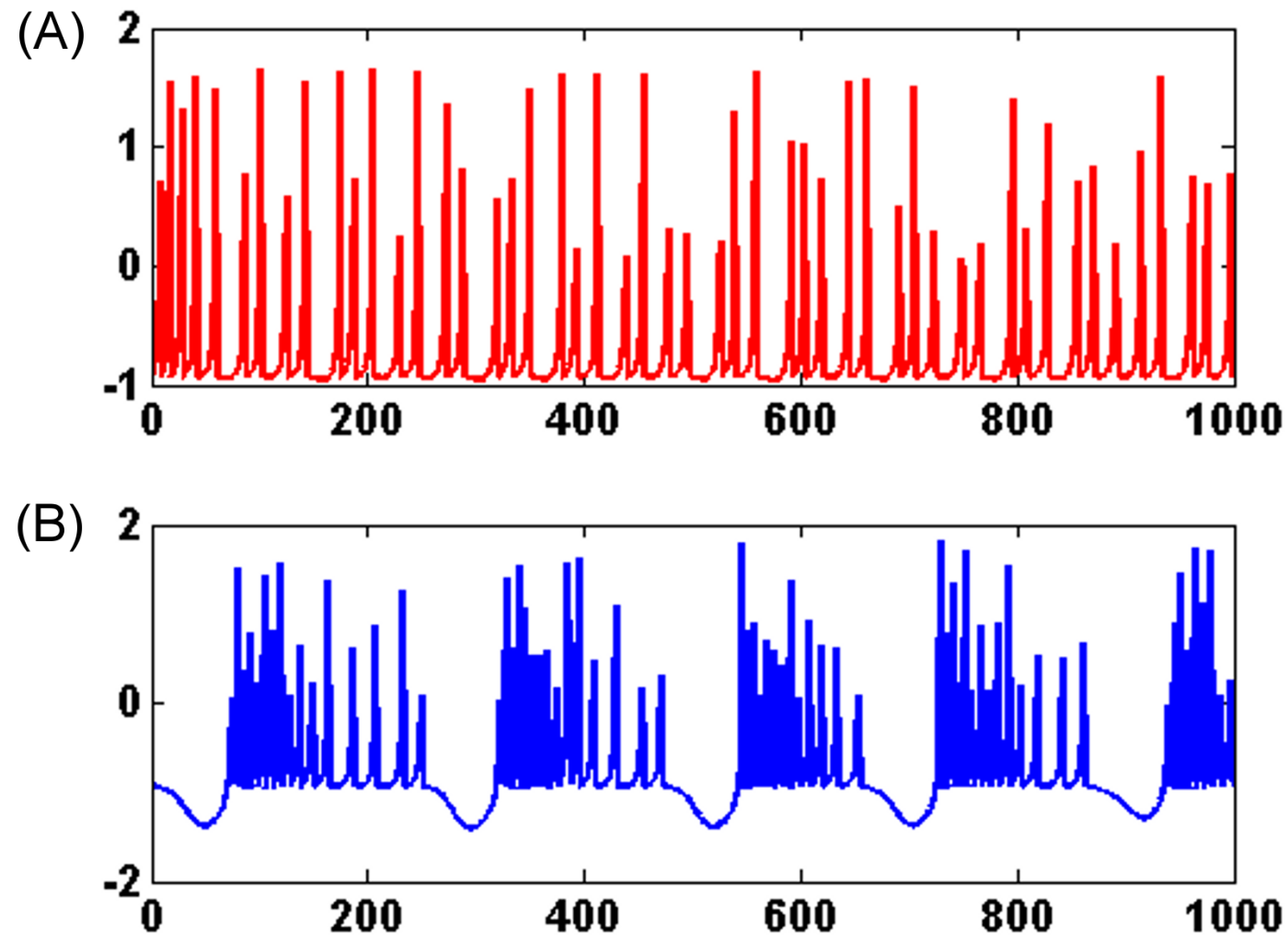

Fig 10: **A single neuron exhibits low and high entropy firing patterns.**
Simulation of a single cortical neuron from the Hindmarsh-Rose neuron model [39] showing two different kinds of behaviour S1 Text. (A) Membrane voltage as a function of time for regular firing exhibited by the neuron when the external current applied is $I = 3.31$. Entropy and Complexities: $H = 0.8342$ bits and $ETC = 0.1910$, $LZ = 0.6879$. (B) Membrane voltage as a function of time for chaotic or irregular spiking exhibited by the neuron when $I = 3.280$. Entropy and Complexities: $H = 0.9295$ bits and $ETC = 0.2211$, $LZ = 0.7262$. Thus, for the same neuron, under two different excitations, the neuron manifests low as well as high entropy behaviour (low and high ETC/LZ complexities correspondingly).

**Table 2:** An exhaustive chronological list of brain complexity measures with their short definitions, theoretical strength, process or capacity, current state dependency, experimental readiness and any other remarks.

| Name | Definition | Tht. Strength | Process/ Capacity | Ct St Dependency | Exp. readiness | Remarks |
|---|---|---|---|---|---|---|
| Neural Complexity [12] (1994) | Sum of average mutual information for all bipartitions of the system. | Strong | Process | Yes | Low | |
| Causal density [13] (2003) | "A measure of causal interactivity that captures dynamical heterogeneity among network elements (differentiation) as well as their global dynamical integration [13]." | Strong | Process | Yes | Low | Calculated by applying "Granger causality". |
| $\Phi$ (IIT 1.0) [10] (2004) | It is the amount of causally effective information that can be integrated across the informational weakest link of a subset of elements. | Medium | Capacity | Yes | Low | Provided the hypothesis for "Information Integrated Theory of Consciousness." Applicable only to stationary systems. |
| $\varphi$ (IIT 2.0 ) [17–19] (2008) | Measure of the information generated by a system when it transitions to one particular state out of a repertoire of possible states, to the extent that this information (generated by the whole system) is over and above the information generated independently by the parts. | Strong | Capacity | Yes | Low | Extension of IIT 1.0 to discrete dynamical systems. |
| $\Phi_E$ and $\Phi_{AR}$ [29] (2011) | Rather than measuring information generated by transitions from a hypothetical maximum entropy past state, $\Phi_E$ instead utilizes the actual distribution of the past state. "$\Phi_{AR}$ can be understood as a measure of the extent to which the present global state of the system predicts the past global state of the system, as compared to predictions based on the most informative decomposition of the system into its component parts [29]." | Strong | Process | No | Medium | $\Phi_E$ is applicable to both discrete and continuous systems with either Markovian or non-Markovian dynamics. $\Phi_{AR}$ is same as $\Phi_E$ for gaussian systems [29]. $\Phi_E$ and $\Phi_{AR}$ fail to satisfy upper and lower bounds of integrated information [22]. However, the authors propose variants of these measures which are well bounded. |
| PCI [7] (2013) | "The normalized Lempel-Ziv complexity of the spatiotemporal pattern of cortical activation triggered by a direct Transcranial Magnetic Stimulation (TMS) perturbation [7]." | Weak | Process | Unknown | High | While PCI proves to be a reasonable objective measure of consciousness in healthy individuals during wakefulness, sleep and anaesthesia, as well as in patients who had emerged from coma, it lacks solid theoretical connections to integrated information theories. |
| $\Phi^{Max}$ (IIT 3.0) [5, 6] (2012-14) | Measure of the Information that is specified by a system that is irreducible to that specified by its parts. "It is calculated as the distance between the conceptual structure specified by the intact system and that specified by its minimum information partition [40]." | Strong | Capacity | Yes | Low | IIT 3.0 introduces major changes over IIT 2.0 and IIT 1.0: (i) considers how mechanisms in a state constrain both the past and the future of a system; (ii) emphasis on "a difference that makes a difference", and not simply "a difference", (iii) Concept has proper metric - Earth Mover's Distance (EMD) [6]. Limitations: Current-state Dependency, Computationally expensive, Inability to handle continuous neurophysiological data. |
| $\psi$ [30] (2014) | $\psi$ is a principled infotheoretic measure of irreducibility to disjoint parts, derived using Partial Information Decomposition (PID), that resides purely within Shannon Information Theory. | Medium | Capacity | No | Low | $\psi$ compares to $\varphi$ (IIT 2.0) instead of $\Phi^{Max}$ (IIT 3.0). Address the three major limitations of $\phi$ in [18]: State-dependency and entropy; issues with duplicate computation and mismatch of the intuition of "cooperation by diverse parts" [30]. Has desirable properties such as not needing a MIP normalization and being substantially faster to compute. |
| $\Phi^*$ [22] (2016) | "It represents the difference between "actual" and "hypothetical" mutual information between the past and present states of the system." It is computed using the idea of mismatched decoding developed from information theory [22]. | Strong | Capacity | Yes | Medium | Emphasis on theoretical requirements: First, the amount of integrated information should not be negative. Second, the amount of integrated information should never exceed information generated by the whole system. Focuses on IIT 2.0, rather IIT 3.0. |
| $\Phi^*_{MMP}$ and $\Phi^{AR}_{MMP}$ [31] (2016) | Introduction of Maximum Modularity Partition (MMP), which is quicker than MIP to compute the integrated information for larger networks. In combination with $\Phi^*$ and $\Phi_{AR}$, MMP yields two new measures $\Phi^*_{MMP}$ and $\Phi^{AR}_{MMP}$. | Strong | Capacity ($\Phi^*_{MMP}$), Process ($\Phi^{AR}_{MMP}$) | Yes ($\Phi^*_{MMP}$), No ($\Phi^{AR}_{MMP}$) | Medium | The new measures are compared with $\Phi^*$, $\Phi_{AR}$ and Causal Density and based on the idea that human brain has modular organisation in its anatomy and functional architecture. Calculating Integrated Information across MMP reflects underlying functional architecture of neural networks. |
| $\Phi^C$ (this paper) | The maximally-aggregate differential normalised Lempel-Ziv (LZ) or normalized Effort-To-Compress (ETC) complexity for the time series data of each node of a network, generated by maximum entropy and zero entropy perturbations of each possible atomic bipartition of an $N$-node network. | High | Both | Low | Medium | Bridges the gap between theoretical and empirical approaches for computing brain complexity. Based on the idea that brain behaves as an integrated system and acknowledging the similarity between compressionism and integrated information, $\Phi^C$ is based on compression-complexity measures and not infotheoretic measures. |

## Concluding Remarks

In summary, we proposed a Compression-Complexity measure of integrated information which incorporates various well supported approaches to determine quantity of integrated information in a network. Some of these approaches adopted by our measure $\Phi^C$ are: using atomic bipartitions, moving beyond MIP approach [31], MEP and ZEP and then recording activity from all the nodes of the network. Furthermore, we have proposed, for the first time, the Differential Compression-Complexity Response Distribution (dCCRD) which can potentially play an important role going forward in understanding the distribution of integrated information in a network.

No doubt, there are more disputed opinions and measures of consciousness now, than ever before. We need to move towards a more theoretically-sound, comprehensive, empirically simplistic and synergistic coalition of different measures which could be applied in the clinic. A combination of different approaches that addresses the interplay between differentiation, integration and entropy is needed. By proposing a compression-complexity based approach, we have taken the first step towards such an end.

# Methods

## Calculation of $\Phi$

We compute $\Phi$ for the following configuration − all possible 3-node networks with logic gates: $XOR$, $OR$, $AND$. The network is fully connected i.e. each node is connected to every other node in the network with a bi-directional connection and no node has any self loop. In this case, there are a total of 10 distinct possible networks and for each 3-node networks there are 8 possible current states of the network.

Using the PyPhi 0.7.0 Python library [6, 21] for computing integrated information, we calculate the values of $\Phi$ for the current state of each network and then calculate the mean of all values ($< \Phi >$). We repeat the same experiment for 4 and 5-node networks. For further details on computing $\Phi$, refer to [6].

## Calculation of $\Phi^C$ $({}_{LZ}\Phi^C$ and ${}_{ETC}\Phi^C)$

To compute the proposed compression-complexity measure, $\Phi^C$, the methods employed are described below.

### Maximum Entropy Perturbation (MEP)

The input to the perturbed node is a maximum entropy time series $\{P_t\}$ which is obtained as follows:

$$\begin{aligned} RAND_t &= rand(0,1) \\ P_t &= 0, \quad \text{if } 0 \leq RAND_t \leq 0.5, \\ &= 1, \quad \text{if } 0.5 < RAND_t \leq 1, \end{aligned}$$

where $rand(0,1)$ generates a uniform random variable between 0 and 1; discrete time $t = 1, 2, \ldots, LEN$, where $LEN$ is the length of the time series generated. We have chosen $LEN = 200$ in our computations.

### Zero Entropy Perturbation (ZEP)

The input to the perturbed node is a zero entropy time series $\{Q_t\} = constant$ of length $LEN = 200$.

### Differential Compression-Complexity Response Distribution (dCCRD)

The perturbation to the $i^{th}$ node is done by independently injecting the MEP and ZEP time series $\{P_t\}$ and $\{Q_t\}$ to node $i$. The two independent sets of output time series $\{T_j^{MEP}\}$ and $\{T_j^{ZEP}\}$ from the remaining $N-1$ nodes (indexed by $j \neq i$) are collected. We compute the differential compression-complexity of the $j^{th}$ time series for the $i^{th}$ perturbed node as follows:

$${}_{\text{ETC}}\varphi^{\text{C}}_{j(i)} = \text{Compute_ETC_Complexity}(T_j^{MEP}) - \text{Compute_ETC_Complexity}(T_j^{ZEP}), \tag{1}$$

where $j = 1, 2, \ldots, N$ and $j \neq i$. Thus, dCCRD for the $i^{th}$ perturbed node is obtained as the following set:

$$dCCRD_{ETC}(i) = \{{}_{\text{ETC}}\varphi^{\text{C}}_{1(i)}, {}_{\text{ETC}}\varphi^{\text{C}}_{2(i)}, \cdots, {}_{\text{ETC}}\varphi^{\text{C}}_{j(i)}, \cdots, {}_{\text{ETC}}\varphi^{\text{C}}_{N(i)}\}, \quad j \neq i.$$

We thus obtain $\{dCCRD_{ETC}(i)\}$ for all perturbed nodes $i = 1, 2, \ldots, N$. The subroutine Compute_ETC_Complexity(·) employs the normalized Effort-To-Compress (ETC) complexity measure, a description of which can be found in S1 Text. ETC uses the lossless compression algorithm called Non-Sequential Recursive Pair Substitution (NSRPS) and it denotes the number of iterations needed for NSRPS to transform the input sequence to a constant sequence. ETC has been found to be more successful as a complexity measure in practical applications (in short and noisy real-world sequences) than infotheoretic measure such as entropy [16, 41].

### Aggregate Differential Compression-Complexity Measure

Once we have the dCCRD for all the perturbed nodes, the aggregate differential compression-complexity measure is obtained as follows:

$$_{\text{ETC}}\phi^{C}_{(i)} = \sum_{j=1}^{j=N} {}_{\text{ETC}}\varphi^{\text{C}}_{j(i)}, \qquad j \neq i,$$

where $i = 1, 2, 3, \ldots, N$.

### Maximal Aggregate Differential Compression-Complexity

We finally obtain:

$$_{\text{ETC}}\Phi^{C} = \max({}_{\text{ETC}}\phi^{C}_{(1)}, {}_{\text{ETC}}\phi^{C}_{(2)}, \cdots, {}_{\text{ETC}}\phi^{C}_{(N)}).$$

For obtaining the other measure $_{LZ}\Phi^{C}$, we replace Compute_ETC_Complexity(·) in Eq 1 with Compute_LZ_Complexity(·). The subscript $LZ$ instead of $ETC$ is carried forward, but the steps remain effectively the same. Compute_LZ_Complexity(·) employs the normalized Lempel-Ziv complexity measure, a description of which can be found in S1 Text.

## Supporting Information

**S1 Text. Supplementary Methods.** Description of methods for Lempel-Ziv complexity, Effort-To-Compress complexity, linear regression of measures of integrated information as a function of entropy of nodes, and Hindmarsh-Rose neuron model.

**S1 Table. Tables of $\Phi$, $_{ETC}\Phi^{C}$ and $_{LZ}\Phi^{C}$ for all 3, 4, 5-node networks.**

**S1 File. An example implementation of $_{ETC}\Phi^{C}$.** 'PhiC_ETC_Fig3.m' shows the step-by-step computation of $_{ETC}\Phi^{C}$ for the example network $ABC$ with 3 nodes, as depicted in Fig 3. 'ETC.m' is the subroutine for the computation of normalized "Effort-To-Compress" (ETC) measure (required to run 'PhiC_ETC_Fig3.m'). 'MEP_TimeSeries.txt' and 'ZEP_TimeSeries.txt' are the text files containing the timeseries for the network $ABC$ when each of its bipartition is perturbed with a random binary sequence and a constant binary sequence (either all zeros or all ones) respectively.

## Acknowledgments

We gratefully acknowledge the help extended by Will Mayner (University of Wisconsin) for assisting with PyPhi Python package. We owe special thanks to Adam Barett for useful feedback and anonymous reviewers' feedback which enabled us to improve our method.

# S1 Text - Supplementary Methods

## Lempel-Ziv Complexity (LZ)

In our study, we have used the Lempel-Ziv complexity measure [1] for computing the compression-complexity of a time series. Lempel-Ziv complexity is a popular measure used in diverse applications. In order to compute the Lempel-Ziv complexity (or LZ) of an input time series, $X = \{x_i\}_{i=1}^{i=n} = x_1 x_2 \ldots x_n$, it is parsed from left to right in order to identify the number of distinct patterns present in $X$. This method of parsing has been proposed in [1] and is related to the universal compression algorithm [2].

We reproduce below a very succinct description of the algorithm for computing LZ complexity, taken from [3]. Let $S = s_1 s_2 \cdots s_n$ denote the input sequence; $S(i,j)$ denote a substring of $S$ that starts at position $i$ and ends at position $j$; $V(S)$ denote the set of all substrings $\{S(i,j), i = 1, 2, \cdots n; j \geq i\}$. For example, let $S = abc$, then $V(S) = a, b, c, ab, bc, abc$. The parsing mechanism involves a left-to-right scan of the symbolic sequence $S$. Start with $i = 1$ and $j = 1$. A substring $S(i,j)$ is compared with all strings in $V(S(i,j-1))$ (Let $V(S(1,0)) = \{\}$, the empty set). If $S(i,j)$ is present in $V(S(1,j-1))$, then increase $j$ by 1 and repeat the process. If the substring is not present, then place a dot after $S(i,j)$ to indicate the end of a new component, set $i = j+1$, increase $j$ by 1, and the process continues. This parsing procedure continues until $j = n$, where $n$ is the length of the symbolic sequence. For example, the sequence '*aacgacga*' is parsed as '*a.ac.g.acga.*'. By convention, a dot is placed after the last element of the symbolic sequence and the number of dots gives us the number of distinct words which is taken as the LZ complexity, denoted by $c(n)$. In this example, the number of distinct words (LZ complexity) is 4. In order to be able to compare the LZ complexity of sequences of different lengths, a normalized measure is proposed [4].

$$C_{LZ} = (c(n)/n) log_{\alpha} n.$$

where $\alpha$ denotes the number of unique symbols in the input time series.

## Effort-To-Compress Complexity (ETC)

Effort-To-Compress (ETC) is a recently proposed complexity measure that measures the effort required by a lossless compression algorithm to compress the input time series/sequence [5]. The lossless compression algorithm known as Non-sequential Recursive Pair Substitution (NSRPS) [6] is used. The algorithm for compressing the input time-series/sequence proceeds as follows. At the first iteration, the pair of symbols which has maximum number of occurrences is replaced by a new symbol. For example, the input sequence '11010010' is transformed into '12202' in the first iteration since the pair '10' has maximum number of occurrences (when compared with the pairs '00', '01' and '11'). In the second iteration, '12202' is transformed to '3202'. The algorithm proceeds in this manner until the length of the transformed string shrinks to 1 or the transformed sequence reduces to a constant sequence. In either cases, the algorithm terminates. For our example, the algorithm transforms the input sequence $11010010 \mapsto 12202 \mapsto 3202 \mapsto 402 \mapsto 52 \mapsto 6$, and thus takes 5 iterations to halt.

The ETC complexity measure is defined as $ETC_{val}$, the number of iterations required for the input sequence to be transformed to a constant sequence through the usage of NSRPS algorithm. This quantity is always a non-negative integer that is bounded between 0 and $L-1$, where $L$ is the length of the input sequence. The normalized version of the measure is given by: $ETC_{norm.} = \frac{ETC_{val}}{L-1}$. Note that $0 \leq \frac{ETC_{val}}{L-1} \leq 1$. For our example, $ETC_{norm.} = \frac{5}{8-1} = \frac{5}{7} = 0.7143$.

## Explanation of the example in Fig. 3

In this section, we describe all the steps for the computation of $_{ETC}\Phi^C$ for the 3-node network of example in Fig. 3, for which the all code files are available in *S1 File*. The network $ABC$ is constituted of three logic gates: $A = OR$, $B = AND$, $C = XOR$ in the current state $(1, 0, 0)$ for which the value of $_{ETC}\Phi^C$ is computed as follows.

- **Step 1: Atomic Bipartitioning** - The network $ABC$ is bipartitioned into 3 possible atomic bipartitions i.e. $A - BC$, $B - AC$ and $C - AB$. $A - BC$ denotes that, when node $A$ is perturbed, the output time series is obtained from the nodes $B$ and $C$. And similarly, biparitions $B - AC$ and $C - AB$ can be defined.

- **Step 2: Zero Entropy Perturbation (ZEP) and Maximum Entropy Perturbation (MEP)** - Each atomic bipartition is independently perturbed with a Maximum Entropy Perturbation (MEP) which is a random input binary time series (length= 200) as well as a Zero Entropy Perturbation (ZEP) which is a constant sequence (length= 200). We take two MEP time series (a random sequence): one which starts with 1 and another which starts with 0. Similarly, we take two ZEP time series (a constant sequence): one which starts with 1, followed by all ones and another which starts with 0, followed by all zeros. This is done so that the appropriate time series could be used according to the current state of a particular node. For this example, in the bipartition $A - BC$, node $A$ is perturbed with a MEP time series starting with 1 and a ZEP time series with all ones, since the current state of node $A$ is 1. Similarly, for the bipartitions $B - AC$ and $C - AB$, nodes $B$ and $C$ are independently perturbed with MEP time series starting with 0 and ZEP time series with all zeros, since the current states of both the nodes $B$ and $C$ is 0. These time series are provided in supplementary material - *S1 File* as *MEP_TimeSeries.txt* and *ZEP_TimeSeries.txt*.

- **Step 3: Differential Compression-Complexity Response Distribution (dCCRD)** - This is computed by taking the difference between

MEP and ZEP complexities for each output time series from the remaining two unperturbed nodes. In this example, for the bipartition $A - BC$, the $\{MEP, ZEP\}$ complexity values for nodes $B$ and $C$ are $\{0.25628, 0.015075\}$ and $\{0.32161, 0.015075\}$ respectively. Thus, the difference: (0.25628 - 0.015075) and (0.32161 − 0.015075) forms the Differential Compression-Complexity Response Distribution (dCCRD) for the bipartition $A - BC$. So, $\{{}_{\mathrm{ETC}}\varphi^{\mathrm{C}}_{\mathrm{B(A)}} = 0.241,\ {}_{\mathrm{ETC}}\varphi^{\mathrm{C}}_{\mathrm{C(A)}} = 0.307\}$ represents the dCCRD of the time series obtained from the nodes $B$ and $C$ respectively, when the node $A$ is perturbed. Similarly, the dCCRD for the other two bipartitions $B - AC$ and $C - AB$ are: $\{{}_{\mathrm{ETC}}\varphi^{\mathrm{C}}_{\mathrm{A(B)}} = 0.241,\ {}_{\mathrm{ETC}}\varphi^{\mathrm{C}}_{\mathrm{C(B)}} = 0.296\}$,$\{{}_{\mathrm{ETC}}\varphi^{\mathrm{C}}_{\mathrm{A(C)}} = 0.186,\ {}_{\mathrm{ETC}}\varphi^{\mathrm{C}}_{\mathrm{B(C)}} = 0.281\}$ respectively.

- **Step 4: Aggregate Differential Compression-Complexity Measure** - The individual values of each dCCRD are summed for each bipartitioned-perturbed network. For this example, in case of bipartition $A - BC$, the dCCRD values of the nodes $B$ (0.241) and $C$ (0.307) are summed up to obtain Aggregate Differential Compression-Complexity Measure for the bipartition $A - BC$ as 0.548. Mathematically, ${}_{\mathrm{ETC}}\phi^{\mathrm{C}}_{\mathrm{(A)}} = {}_{\mathrm{ETC}}\ \varphi^{\mathrm{C}}_{\mathrm{B(A)}} + {}_{\mathrm{ETC}}\ \varphi^{\mathrm{C}}_{\mathrm{C(A)}}$ and similarly, ${}_{\mathrm{ETC}}\phi^{\mathrm{C}}_{\mathrm{(B)}}$ and ${}_{\mathrm{ETC}}\phi^{\mathrm{C}}_{\mathrm{(C)}}$ can be computed. Therefore, $\mathbf{{}_{ETC}\phi^{C}_{(A)}} = 0.548$, $\mathbf{{}_{ETC}\phi^{C}_{(B)}} = 0.538$ and $\mathbf{{}_{ETC}\phi^{C}_{(C)}} = 0.467$.
- **Step 5: Maximal-Aggregate Differential Compression-Complexity** - It is represented as ${}_{ETC}\Phi^{C}$, which is nothing but the maximum of the Aggregate Differential Compression-Complexity measures. In this example, the maximum value among the aggregate dCCRDs of each biparition is 0.548. Mathematically, ${}_{\mathrm{ETC}}\Phi^{\mathrm{C}} = \max({}_{\mathrm{ETC}}\phi^{\mathrm{C}}_{\mathrm{(A)}}, {}_{\mathrm{ETC}}\ \phi^{\mathrm{C}}_{\mathrm{(B)}}, {}_{\mathrm{ETC}}\ \phi^{\mathrm{C}}_{\mathrm{(C)}})$. Thus, $\mathbf{{}_{ETC}\Phi^{C}} = 0.548$.

## Linear regression of measures of integrated information as a function of entropy of nodes

Let $Y$ denote measures of integrated information discussed in our study. Thus, $Y$ could be any of $< \Phi >$, $<_{LZ}\ \Phi^C >$, or $<_{ETC}\ \Phi^C >$. We shall perform a linear regression (least squares) between the dependent variable $Y$ and the explanatory (independent) variables 'entropy' of the nodes and the 'number of nodes'. We have considered three different kinds of logic gates $XOR$, $AND$ and $OR$. The output of $XOR$ gate has higher entropy ($H = 1$ bit) than $AND$ and $OR$ gates ($H = 0.8113$ bits). The independent variables are the two types of nodes - high entropy nodes, $n_{high}$ of them each with $H_{high}$, and low entropy nodes, $n_{low}$ of them each with $H_{low}$. We seek to fit the following function:

$$\begin{aligned} Y &= f(n_{high}, H_{high}, n_{low}, H_{low}), \\ &= n_{high}H_{high}x_{high} + n_{low}H_{low}x_{low}, \end{aligned}$$

where we are required to determine the unknown coefficients $x_{high}$ and $x_{low}$.

**An example**

As an example, let us consider all 3-node networks and obtain a linear fit between $Y = <\Phi>$ and the independent variables $n_{high}H_{high}$ and $n_{low}H_{low}$. The relevant values are given in Table 1. Also, note that $H_{high} = 1$ bit and $H_{low} = 0.8113$ bits.

Table 1: **The values of $Y = <\Phi>$ for all 3 node networks and the number of high entropy ($n_{high}$) and low entropy gates ($n_{low}$), as well as the predicted output $\hat{Y}$ from linear regression.**

| Networks | $Y$ | $n_{high}$ | $n_{low}$ | $\hat{Y}$ |
|---|---|---|---|---|
| XOR-XOR-XOR | 3.000 | 3 | 0 | 3.3301 |
| XOR-XOR-OR | 2.568 | 2 | 1 | 2.3343 |
| XOR-XOR-AND | 2.568 | 2 | 1 | 2.3343 |
| OR-OR-XOR | 1.704 | 1 | 2 | 1.3385 |
| AND-AND-XOR | 1.422 | 1 | 2 | 1.3385 |
| OR-AND-XOR | 0.946 | 1 | 2 | 1.3385 |
| AND-AND-OR | 0.312 | 0 | 3 | 0.3427 |
| OR-OR-AND | 0.312 | 0 | 3 | 0.3427 |
| AND-AND-AND | 0.277 | 0 | 3 | 0.3427 |
| OR-OR-OR | 0.277 | 0 | 3 | 0.3427 |

A linear regression (least-squares) is performed between the dependent variable $Y$ and the explanatory/independent variables $n_{high}H_{high}$ and $n_{low}H_{low}$. The predicted output $\hat{Y}$ displayed above shows that it is quite close to $Y$.

For the above example, we obtain the least squares solution as $\hat{x}_{high} = 1.11$ and $\hat{x}_{low} = 0.1408$. The predicted value of $Y$ is given by

$$\hat{Y} = n_{high}H_{high}\hat{x}_{high} + n_{low}H_{low}\hat{x}_{low}.$$

## Hindmarsh-Rose Neuron Model

The equations of the Hindmarsh-Rose neuron model [7] in dimensionless form are:

$$\begin{aligned}
\dot{S} &= P + 3S^2 - S^3 - Q + I, \\
\dot{P} &= 1 - 5S^2 - P, \\
\dot{Q} &= -r\big[Q - 4(S + \frac{8}{5})\big],
\end{aligned}$$

where $S(t)$ is the membrane voltage of a single neuron. The model has the following control parameters: $I$ and $r$, where the former is the external current applied and the later is the internal state of the neuron. In our simulations we have chosen $r = 0.0021$. The values of $I$ chosen are $I = 3.310$ for simulating regular spiking and $I = 3.28$ for simulating irregular/chaotic spiking. We have used a window of length 2 and if the value of $S(t)$ exceeded a threshold of $-0.1$ in this window, we count it as a spike ('1'). The resulting sequence of 0s (no-spike) and 1s (spike) is used for computing Shanon entropy, LZ and ETC complexities.

# S1 Table: Tables of integrated information for all 3, 4, 5 node networks.

**Table 1.** $< \Phi >$ **values (with standard deviations) in decreasing order for (a) 3, (b) 4, (c) 5 node networks.**

(a)

| Network No. | Networks | $< \Phi > \pm Stdev.$ |
|---|---|---|
| 1 | $XOR - XOR - XOR$ | $3 \pm 1.203$ |
| 2 | $XOR - XOR - OR$ | $2.568 \pm 1.312$ |
| 3 | $XOR - XOR - AND$ | $2.568 \pm 1.312$ |
| 4 | $OR - OR - XOR$ | $1.704 \pm 1.680$ |
| 5 | $AND - AND - XOR$ | $1.422 \pm 1.434$ |
| 6 | $OR - AND - XOR$ | $0.946 \pm 0.636$ |
| 7 | $AND - AND - OR$ | $0.312 \pm 0.091$ |
| 8 | $OR - OR - AND$ | $0.312 \pm 0.091$ |
| 9 | $AND - AND - AND$ | $0.277 \pm 0.093$ |
| 10 | $OR - OR - OR$ | $0.277 \pm 0.093$ |

(b)

| Network No. | Networks | $< \Phi > \pm Stdev.$ |
|---|---|---|
| 1 | $XOR - XOR - XOR - XOR$ | $5.5 \pm 0$ |
| 2 | $XOR - XOR - XOR - OR$ | $2.793 \pm 2.062$ |
| 3 | $XOR - XOR - XOR - AND$ | $2.793 \pm 2.062$ |
| 4 | $OR - OR - XOR - XOR$ | $2.083 \pm 1.735$ |
| 5 | $AND - AND - XOR - XOR$ | $2.083 \pm 1.735$ |
| 6 | $OR - OR - OR - XOR$ | $1.183 \pm 1.037$ |
| 7 | $AND - AND - AND - XOR$ | $1.183 \pm 1.037$ |
| 8 | $XOR - XOR - AND - OR$ | $0.826 \pm 1.116$ |
| 9 | $AND - AND - OR - XOR$ | $0.325 \pm 0.218$ |
| 10 | $OR - OR - AND - XOR$ | $0.325 \pm 0.218$ |
| 11 | $AND - AND - AND - OR$ | $0.126 \pm 0.057$ |
| 12 | $OR - OR - OR - AND$ | $0.126 \pm 0.057$ |
| 13 | $AND - AND - OR - OR$ | $0.119 \pm 0.053$ |
| 14 | $OR - OR - OR - OR$ | $0.092 \pm 0.051$ |
| 15 | $AND - AND - AND - AND$ | $0.092 \pm 0.051$ |

(c)

| Network No. | Networks | $< \Phi > \pm Stdev.$ |
|---|---|---|
| 1 | $XOR - XOR - XOR - XOR - XOR$ | $7.031 \pm 1.095$ |
| 2 | $AND - AND - AND - AND - XOR$ | $0.988 \pm 1.997$ |
| 3 | $XOR - XOR - XOR - OR - XOR$ | $0.864 \pm 1.997$ |
| 4 | $XOR - XOR - XOR - AND - XOR$ | $0.864 \pm 1.997$ |
| 5 | $XOR - XOR - XOR - OR - AND$ | $0.748 \pm 1.575$ |
| 6 | $AND - AND - AND - XOR - XOR$ | $0.491 \pm 1.087$ |
| 7 | $XOR - XOR - XOR - AND - AND$ | $0.426 \pm 0.969$ |
| 8 | $XOR - XOR - XOR - OR - OR$ | $0.406 \pm 0.815$ |
| 9 | $OR - OR - AND - XOR - XOR$ | $0.364 \pm 0.623$ |
| 10 | $AND - AND - OR - XOR - XOR$ | $0.297 \pm 0.387$ |
| 11 | $OR - OR - OR - XOR - XOR$ | $0.262 \pm 0.440$ |
| 12 | $OR - OR - OR - AND - XOR$ | $0.175 \pm 0.117$ |
| 13 | $AND - AND - OR - OR - XOR$ | $0.157 \pm 0.066$ |
| 14 | $AND - AND - AND - OR - XOR$ | $0.154 \pm 0.054$ |
| 15 | $OR - OR - OR - OR - XOR$ | $0.117 \pm 0.151$ |
| 16 | $OR - OR - OR - OR - AND$ | $0.047 \pm 0.027$ |
| 17 | $AND - AND - AND - OR - AND$ | $0.047 \pm 0.027$ |
| 18 | $AND - AND - AND - OR - OR$ | $0.046 \pm 0.025$ |
| 19 | $OR - OR - OR - AND - AND$ | $0.046 \pm 0.025$ |
| 20 | $AND - AND - AND - AND - AND$ | $0.028 \pm 0.022$ |
| 21 | $OR - OR - OR - OR - OR$ | $0.028 \pm 0.022$ |

There are a total of 10, 15 and 21 networks with 3, 4 and 5 nodes respectively, composed of three logic gates $AND$, $OR$, $XOR$. The above tables shows the hierarchy of these networks with respect to their $< \Phi >$ values.

**Table 2.** $<_{\mathrm{ETC}}\ \Phi^{\mathrm{C}}>$ **values (with standard deviations) in decreasing order for (a) 3, (b) 4, (c) 5 node networks.**

(a)

| Network No. | Networks | $<_{\mathrm{ETC}}\ \Phi^{\mathrm{C}}> \pm Stdev$ |
|---|---|---|
| 1 | $XOR-XOR-XOR$ | 0.604 ± 0.020 |
| 3 | $XOR-XOR-AND$ | 0.603 ± 0.022 |
| 2 | $XOR-XOR-OR$ | 0.603 ± 0.022 |
| 4 | $OR-OR-XOR$ | 0.540 ± 0.024 |
| 5 | $AND-AND-XOR$ | 0.525 ± 0.036 |
| 6 | $OR-AND-XOR$ | 0.517 ± 0.062 |
| 8 | $OR-OR-AND$ | 0.465 ± 0.054 |
| 7 | $AND-AND-OR$ | 0.465 ± 0.055 |
| 9 | $AND-AND-AND$ | 0.060 ± 0.064 |
| 10 | $OR-OR-OR$ | 0.060 ± 0.064 |

(b)

| Network No. | Networks | $<_{\mathrm{ETC}}\ \Phi^{\mathrm{C}}> \pm Stdev$ |
|---|---|---|
| 1 | $XOR-XOR-XOR-XOR$ | 0.840 ± 0.103 |
| 2 | $XOR-XOR-XOR-OR$ | 0.807 ± 0.101 |
| 3 | $XOR-XOR-XOR-AND$ | 0.807 ± 0.101 |
| 8 | $XOR-XOR-AND-OR$ | 0.765 ± 0.07 |
| 4 | $OR-OR-XOR-XOR$ | 0.730 ± 0.09 |
| 5 | $AND-AND-XOR-XOR$ | 0.724 ± 0.09 |
| 9 | $AND-AND-OR-XOR$ | 0.721 ± 0.069 |
| 10 | $OR-OR-AND-XOR$ | 0.720 ± 0.067 |
| 11 | $AND-AND-AND-OR$ | 0.638 ± 0.066 |
| 12 | $OR-OR-OR-AND$ | 0.638 ± 0.067 |
| 7 | $AND-AND-AND-XOR$ | 0.613 ± 0.066 |
| 6 | $OR-OR-OR-XOR$ | 0.611 ± 0.066 |
| 13 | $AND-AND-OR-OR$ | 0.318 ± 0.088 |
| 15 | $AND-AND-AND-AND$ | 0.017 ± 0.050 |
| 14 | $OR-OR-OR-OR$ | 0.017 ± 0.050 |

(c)

| Network No. | Networks | $<_{\mathrm{ETC}}\ \Phi^{\mathrm{C}}> \pm Stdev$ |
|---|---|---|
| 1 | $XOR-XOR-XOR-XOR-XOR$ | 1.206 ± 0.041 |
| 4 | $XOR-XOR-XOR-AND-XOR$ | 1.204 ± 0.043 |
| 3 | $XOR-XOR-XOR-OR-XOR$ | 1.204 ± 0.044 |
| 5 | $XOR-XOR-XOR-OR-AND$ | 0.935 ± 0.129 |
| 7 | $XOR-XOR-XOR-AND-AND$ | 0.875 ± 0.155 |
| 8 | $XOR-XOR-XOR-OR-OR$ | 0.861 ± 0.184 |
| 10 | $AND-AND-OR-XOR-XOR$ | 0.757 ± 0.168 |
| 9 | $OR-OR-AND-XOR-XOR$ | 0.755 ± 0.165 |
| 11 | $OR-OR-OR-XOR-XOR$ | 0.609 ± 0.044 |
| 6 | $AND-AND-AND-XOR-XOR$ | 0.601 ± 0.023 |
| 13 | $AND-AND-OR-OR-XOR$ | 0.533 ± 0.038 |
| 12 | $OR-OR-OR-AND-XOR$ | 0.522 ± 0.050 |
| 16 | $OR-OR-OR-OR-AND$ | 0.305 ± 0.053 |
| 19 | $OR-OR-OR-AND-AND$ | 0.281 ± 0.063 |
| 15 | $OR-OR-OR-OR-XOR$ | 0.279 ± 0.063 |
| 17 | $AND-AND-AND-OR-AND$ | 0.271 ± 0.035 |
| 2 | $AND-AND-AND-AND-XOR$ | 0.263 ± 0.030 |
| 18 | $AND-AND-AND-OR-OR$ | 0.023 ± 0.048 |
| 21 | $OR-OR-OR-OR-OR$ | 0.016 ± 0.064 |
| 14 | $AND-AND-AND-OR-XOR$ | 0 ± 0 |
| 20 | $AND-AND-AND-AND-AND$ | 0 ± 0 |

There are a total of 10, 15 and 21 networks with 3, 4 and 5 nodes respectively, composed of three logic gates $AND$, $OR$, $XOR$. The above tables shows the hierarchy of these networks with respect to their $<_{\mathrm{ETC}}\ \Phi^{\mathrm{C}}>$ values.

**Table 3.** $<_{\mathrm{LZ}} \Phi^{\mathrm{C}} >$ **(along with standard deviations) in decreasing order for (a) 3, (b) 4, (c) 5 node networks.**

(a)

| Network No. | Networks | $<_{\mathrm{LZ}} \Phi^{\mathrm{C}} > \pm$ *Stdev* |
|---|---|---|
| 1 | $XOR - XOR - XOR$ | 2.026 ± 0.115 |
| 3 | $XOR - XOR - AND$ | 2.026 ± 0.115 |
| 2 | $XOR - XOR - OR$ | 2.026 ± 0.115 |
| 4 | $OR - OR - XOR$ | 1.725 ± 0.040 |
| 6 | $OR - AND - XOR$ | 1.710 ± 0.046 |
| 5 | $AND - AND - XOR$ | 1.658 ± 0.038 |
| 7 | $AND - AND - OR$ | 1.610 ± 0.023 |
| 8 | $OR - OR - AND$ | 1.610 ± 0.023 |
| 9 | $AND - AND - AND$ | 0.038 ± 0.038 |
| 10 | $OR - OR - OR$ | 0.038 ± 0.038 |

(b)

| Network No. | Networks | $<_{\mathrm{LZ}} \Phi^{\mathrm{C}} > \pm$ *Stdev* |
|---|---|---|
| 1 | $XOR - XOR - XOR - XOR$ | 3.029 ± 0.065 |
| 2 | $XOR - XOR - XOR - OR$ | 2.924 ± 0.059 |
| 3 | $XOR - XOR - XOR - AND$ | 2.924 ± 0.059 |
| 5 | $AND - AND - XOR - XOR$ | 2.683 ± 0.102 |
| 4 | $OR - OR - XOR - XOR$ | 2.683 ± 0.102 |
| 8 | $XOR - XOR - AND - OR$ | 2.582 ± 0.175 |
| 9 | $AND - AND - OR - XOR$ | 2.444 ± 0.073 |
| 10 | $OR - OR - AND - XOR$ | 2.444 ± 0.073 |
| 11 | $AND - AND - AND - OR$ | 2.081 ± 0.042 |
| 12 | $OR - OR - OR - AND$ | 2.081 ± 0.042 |
| 7 | $AND - AND - AND - XOR$ | 1.992 ± 0.054 |
| 6 | $OR - OR - OR - XOR$ | 1.992 ± 0.054 |
| 13 | $AND - AND - OR - OR$ | 1.020 ± 0.050 |
| 15 | $AND - AND - AND - AND$ | 0.017 ± 0.034 |
| 14 | $OR - OR - OR - OR$ | 0.017 ± 0.034 |

(c)

| Network No. | Networks | $<_{\mathrm{LZ}} \Phi^{\mathrm{C}} > \pm$ *Stdev* |
|---|---|---|
| 3 | $XOR - XOR - XOR - OR - XOR$ | 4.051 ± 0.233 |
| 4 | $XOR - XOR - XOR - AND - XOR$ | 4.051 ± 0.233 |
| 1 | $XOR - XOR - XOR - XOR - XOR$ | 4.051 ± 0.233 |
| 5 | $XOR - XOR - XOR - OR - AND$ | 3.246 ± 0.264 |
| 7 | $XOR - XOR - XOR - AND - AND$ | 3.103 ± 0.322 |
| 8 | $XOR - XOR - XOR - OR - OR$ | 3.077 ± 0.429 |
| 10 | $AND - AND - OR - XOR - XOR$ | 2.616 ± 0.558 |
| 9 | $OR - OR - AND - XOR - XOR$ | 2.611 ± 0.551 |
| 11 | $OR - OR - OR - XOR - XOR$ | 2.088 ± 0.081 |
| 6 | $AND - AND - AND - XOR - XOR$ | 2.083 ± 0.075 |
| 13 | $AND - AND - OR - OR - XOR$ | 1.717 ± 0.061 |
| 12 | $OR - OR - OR - AND - XOR$ | 1.629 ± 0.030 |
| 19 | $OR - OR - OR - AND - AND$ | 0.984 ± 0.038 |
| 17 | $AND - AND - AND - OR - AND$ | 0.975 ± 0.019 |
| 2 | $AND - AND - AND - AND - XOR$ | 0.972 ± 0.024 |
| 16 | $OR - OR - OR - OR - AND$ | 0.967 ± 0.038 |
| 15 | $OR - OR - OR - OR - XOR$ | 0.967 ± 0.054 |
| 18 | $AND - AND - AND - OR - OR$ | 0.014 ± 0.030 |
| 21 | $OR - OR - OR - OR - OR$ | 0.010 ± 0.038 |
| 14 | $AND - AND - AND - OR - XOR$ | 0 ± 0 |
| 20 | $AND - AND - AND - AND - AND$ | 0 ± 0 |

There are a total of 10, 15 and 21 networks with 3, 4 and 5 nodes respectively, composed of three logic gates $AND$, $OR$, $XOR$. The above tables shows the hierarchy of these networks with respect to their $<_{\mathrm{LZ}} \Phi^{\mathrm{C}} >$ values.